\documentclass[11pt]{article}
\usepackage{geometry}
\usepackage{amsmath}
\usepackage{hyperref}
\usepackage{amssymb}
\usepackage{mathtools}
\usepackage{subcaption}
\usepackage{graphicx}
\usepackage{natbib}
\usepackage{multicol}
\usepackage{algorithm,algpseudocode}

\usepackage{authblk}
\usepackage{enumitem}
\usepackage{amsmath}
\usepackage{hyperref}
\usepackage{tabularx}
\usepackage{xcolor}
\usepackage{amssymb}
\usepackage{float}
\restylefloat{table}
\usepackage{comment}
\usepackage{mathtools}
\usepackage{bm}

\usepackage{algorithm}
\newcommand{\bX}{\bm{X}}

\newcommand{\bZ}{\bm{Z}}

\newcommand{\bU}{\bm{U}}
\newcommand{\bS}{\bm{S}}
\newcommand{\bI}{\bm{I}}

\newcommand{\bC}{\bm{C}}

\newcommand{\bP}{\bm{P}}
\newcommand{\bJ}{\bm{J}}
\newcommand{\bK}{\bm{K}}
\newcommand{\bA}{\bm{A}}
\newcommand{\bW}{\bm{W}}

\newcommand{\bv}{\bm{v}}
\newcommand{\ba}{\bm{a}}

\newcommand{\be}{\bm{e}}
\newcommand{\bF}{\bm{F}}

\newcommand{\bsig}{\bm{\Sigma}}
\newcommand{\bu}{\bm{u}}

\newcommand{\beps}{\bm{\epsilon}}

\newcommand{\bphi}{\bm{\phi}}

\newcommand{\E}{\mathbb{E}}
\newcommand{\blam}{\bm{\Lambda}}
\newcommand{\btheta}{\bm{\theta}}
\newcommand{\bomega}{\bm{\Omega}}

\newcommand{\Var}{\mathrm{Var}}

\newcommand{\bL}{\bm{L}}
\newcommand{\br}{\bm{r}}
\newcommand{\bN}{\bm{N}}

\DeclareMathOperator*{\argmin}{arg\,min}
\DeclareMathOperator*{\argmax}{arg\,max}
\usepackage{graphicx}
\usepackage{float}
\usepackage{wrapfig}
\usepackage{lscape}
\usepackage{arydshln}
\usepackage{multirow}
\usepackage{tikz}
\usepackage{mathtools}

\date{}
\begin{document}

\title{sparseDFM: An R Package to Estimate Dynamic Factor Models with Sparse Loadings}


\author{Luke Mosley}

\author{Tak-Shing T.~Chan}

\author{Alex Gibberd\footnote{Corresponding author (e-mail: a.gibberd@lancaster.ac.uk)}}

\affil{Department of Mathematics and Statistics, Lancaster University, Lancaster, LA1 4YW, United Kingdom}

\maketitle

\abstract{\noindent
  \textbf{sparseDFM} is an R package for the implementation of popular estimation methods for dynamic factor models (DFMs) including the novel Sparse DFM approach of \cite{mosleySAC}. The Sparse DFM ameliorates interpretability issues of factor structure in classic DFMs by constraining the loading matrices to have few non-zero entries (i.e. are sparse). \cite{mosleySAC} construct an efficient expectation maximisation (EM) algorithm to enable estimation of model parameters using a regularised quasi-maximum likelihood. We provide detail on the estimation strategy in this paper and show how we implement this in a computationally efficient way. We then provide two real-data case studies to act as tutorials on how one may use the \textbf{sparseDFM} package. The first case study focuses on summarising the structure of a small subset of quarterly CPI (consumer price inflation) index data for the UK, while the second applies the package onto a large-scale set of monthly time series for the purpose of nowcasting nine of the main trade commodities the UK exports worldwide.}

\section[Introduction]{Introduction}
Dynamic factor models (DFMs) are one of, if not the most, popular tool for summarising the sources of variation among large collections of time series variables. Originally formalised by \cite{geweke1977dynamic}, it was the application of \cite{sargent1977business} showing how just two dynamic factors were able to explain the majority of variance in headline US macroeconomic variables that has led to DFMs popularity in modern day forecasting and structural analysis with `big data'. Their natural state-space representation makes it possible to conduct inference based on Kalman filtering and smoothing techniques and so we obtain a compact framework to handle mixed-frequency and missing data  issues in economic data sets and the ability to
update forecasts in real-time. DFMs have been intensively used in many contexts, including macroeconomic nowcasting \citep{giannone2008nowcasting,banbura2010nowcasting,foroni2014comparison}, building indicators of economic activity \citep{mariano2010coincident, grassi2015euromind}, counterfactual analysis \citep{harvey1996intervention, luciani2015monetary}, and areas away from economics such as psychology \citep{molenaar1985dynamic, fisher2015toward}, the energy sector \citep{wu2013new, lee2016load} and many more. See \cite{stock2011dynamic} and \cite{poncela2021factor} for detailed surveys of the literature. 

The state-space framework of a DFM consists of a measurement equation: 
\begin{equation}
    \bX_t = \blam \bF_t + \beps_t \, ,
    \label{eq:measurement}
\end{equation}
along with a state equation: 
\begin{equation}
    \bF_t = \bA \bF_{t-1} + \bu_t \, ,
    \label{eq:state}
\end{equation}
for observations $t=1,\dots,n$, which together assume that the common dynamics of a large number of stationary time series variables $\bX_t = (X_{1,t}, \dots, X_{p,t})^\top$ are driven by a small number $r \ll p$ of latent factors $\bF_t = (F_{1,t},\dots,F_{r,t})^\top$ which evolve dynamically as a VAR(1) process over time. The $p \times r$ matrix $\blam$ contains factor loadings representing the relationship between the measurements and the underlying latent factors that drive their behaviour. The factor loadings play a critical role in the analysis of DFMs, providing an insight into the underlying structure of the data and the strength of correlation between factors and variables. As we will see, factor loadings are the key parameters in the Sparse DFM extension.

The measurement equation (\ref{eq:measurement}) can be decomposed into a common component $\bm{\chi}_t = \blam\bF_t$ and an idiosyncratic component $\beps_t = (\epsilon_{1,t},\dots,\epsilon_{p,t})^\top$. The common component captures the variation in the time series variables that can be attributed to the underlying latent factors. In a classic DFM, the factors are linear combinations of all variables and therefore $\bm{\chi}_t$ reflects the common sources of variation amongst all variables. The idiosyncratic errors capture the sources of variation that cannot be explained by the factors and are specific to individual series, such as measurement errors. Interpreting the common and idiosyncratic components is important when trying to understand data. For example, in an econometric application, it might be that the common component represents an underlying business cycle, while the idiosyncratic component may represent shocks specific to individual industries or regions \citep{kose2008understanding,forni2021policy}.

To avoid the proliferation of parameters, it is generally assumed that the idiosyncratic errors are zero-mean and cross-sectionally uncorrelated, meaning their covariance matrix, which we denote $\bsig_{\beps}$, is diagonal - this is termed an exact DFM. Under this assumption, the correlation of one series with another occurs only through the latent factors $\bF_t$. When this assumption is relaxed, and the idiosyncratic errors are allowed to be (weakly) cross-correlated, termed an approximate DFM, consistent estimation of the factors is possible only in a high-dimensional setting, i.e. when $p \rightarrow \infty$. Therefore, the `curse of dimensionality' problem, often a burden for analysing time series models, is beneficial when estimating approximate DFMs. In the \textbf{sparseDFM} package we consider two distributions for the idiosyncratic errors: $\beps_t$ is serially uncorrelated IID white noise, or $\beps_t$ follow an AR(1) process.

Despite the many qualities DFMs possess in handling high-dimensional data, dealing with arbitrary patterns of missing data and being able to update predictions in real time, one may argue they suffer from a weakness - a lack of interpretability. 
In the classic DFM, every variable in $\bX_t$ is loaded onto the factors $\bF_t$ via the factor loadings, i.e. $\blam$ is a dense matrix. Often these loading weights are quite large and it is very difficult, if not impossible, to understand how individual series, or groups of series, are influencing the dynamics of the estimated latent factors. If instead, we could obtain factor estimates that are linear combinations of small and distinct groups of variables, where we could associate a given factor estimate with an underlying feature of the observed data, then this would allow us to make statements like: ‘this factor is largely associated with data coming from
this group of variables’. This would be highly informative for users of such data.

A natural way we could impose this kind of interpretability into the factor structure would be through the factor loadings $\blam$. A classic approach to accomplish this is based on factor rotations, which has been widely used to minimise the complexity in the factor loadings to make the structure simpler to interpret. This is useful when you do not have, or would not like to use prior knowledge on factor structure. See \cite{kaiser1958varimax} for the well-established varimax rotation and see \cite{carroll1953analytical} and \cite{jennrich1966rotation} for the well-established quartmin rotation. For a recent paper on varimax rotation see \cite{rohe2020vintage}. The well-established nowcasting paper of \cite{banbura2010nowcasting} use prior knowledge of the variables to partition the factors into mutually independent global, real and nominal factors, assuming that the global factor is loaded by all the variables, while real and nominal factors are specific to real and nominal variables. They show this is helpful when calculating a global factor. \cite{andreou2019inference} is a similar more recent application for grouped factor structure. 

An alternative approach is to utilise modern tools from sparse modelling and regularised estimation to constrain the loading matrix to only have a few non-zero loadings for each factor, i.e. is sparse. Sparse principal components analysis (SPCA) from \cite{zou2006sparse} is a variation of traditional PCA that adds an elastic net penalty (i.e. a mix of $\ell_1$ and $\ell_2$ norms) to induce loadings that are sparse. The interpretable principal components can be viewed as static factors in a factor model and used alongside the Kalman smoother for prediction purposes. See  \cite{croux2011sparse} for a typical macroeconomic forecasting setting where they consider a robustified version. \cite{kristensen2017diffusion} use SPCA to estimate diffusion indexes with sparse loadings. \cite{despois2022identifying} prove that SPCA consistently estimates the factors in an approximate factor model if the $\ell_1$ penalty is of $\mathcal{O}(1/\sqrt{p})$. They compare SPCA with factor rotation methods and show an improved performance when the true loadings structure is sparse.

A new paper by \cite{mosleySAC} implements regularisation (via the $\ell_1$ norm) of factor loadings within an expectation maximisation (EM) algorithm framework allowing the ability to robustly handle arbitrary patterns of missing data, model temporal dependence in the process (i.e. obtain dynamic factors rather than static), and impose weakly informative (sparse) prior knowledge on the factor loadings. Numeric results convey the importance of an EM framework when recovering the true factor loadings and producing accurate forecasts. When the autocorrelation between factors is moderately persistent, they show an improved performance over SPCA for loading recovery and improved performance over a sparse VAR model in forecasting. 

The aim of the package \textbf{SparseDFM} is to offer the tools for the R user to implement dynamic factor models with the option to induce sparse loadings. In summary we offer three different popular DFM estimation techniques plus the novel Sparse DFM estimation technique of \cite{mosleySAC}. We call the estimation options:
\begin{itemize}
    \item \verb+PCA+ - principal components analysis (PCA) for static factors seen in \cite{stock2002forecasting}. 
    \item \verb+2Stage+ - the two-stage framework of PCA plus Kalman filter \& smoother seen in \cite{giannone2008nowcasting} and \cite{doz2011two}.
    \item \verb+EM+ - the quasi-maximum likelihood approach using the EM algorithm to handle arbitrary patterns of missing data seen in \cite{banbura2014maximum}.
    \item \verb+EM-sparse+ - the novel sparse EM approach allowing for sparse factor loadings seen in \cite{mosleySAC}.
\end{itemize}
We allow the user the option of two different idiosyncratic error processes:
\begin{itemize}
    \item \verb+IID+ - errors are IID white noise: $\epsilon_{i,t} \sim N(0, \sigma_i^2)$.
    \item \verb+AR1+ - errors follow an AR(1) process: $\epsilon_{i,t} = \phi_i \epsilon_{i,t-1} + e_{i,t}$ with $e_{i,t} \sim N(0,\sigma_i^2)$.
\end{itemize}
We also allow two different options for estimating the Kalman filter and smoother equations:
\begin{itemize}
    \item \verb+multivariate+ - classic Kalman filter and smoother equations seen in  \cite{shumway1982approach}.
    \item \verb+univariate+ - univariate treatment (sequential processing) of the multivariate equations for fast Kalman filter and smoother seen in \cite{koopman2000fast}.
\end{itemize}

Alternative software that implement classic DFMs in R include the \textbf{MARSS} package of 
\cite{holmes2012marss} which allow for more general state space structures. The nowcasting packages include \textbf{nowcasting} by \cite{de2019nowcasting} and \textbf{nowcastDFM} by \cite{dhopp2021}, that allow mixed-frequency time series nowcasting\footnote{These nowcasting packages have now been removed from the CRAN repository. However, they are still accessible via GitHub.}. Recently, the package \textbf{dfms} by \cite{krantz2022} has been uploaded to CRAN which implement the regular DFM methods with IID errors. They make use of C++ code and therefore is computationally faster than the previous three packages whose implementations are solely in R. Our package \textbf{SparseDFM} is novel in multiple ways. We implement the general DFM estimation methods found in the above packages also using C++ code for speed. We implement the univariate treatment of the Kalman filter/smoother equations of \cite{koopman2000fast} which is computationally much faster than the classic multivariate approach when $p$ is large. We consider AR(1) idiosyncratic errors as well as IID. Finally, we allow the ability to estimate a sparse DFM using an efficient sparsified EM algorithm framework. 

This article is structured as follows. Section \ref{sec2} provides information on the 4 estimation methods we implement in this package: PCA in Section \ref{sec2.1}, 2Stage in Section \ref{sec2.2}, EM in Section \ref{sec2.3} and EM-sparse in \ref{sec2.4}. We provide derivations of the Kalman filter and smoother equations and parameters estimated in the EM algorithm. In the EM-sparse section we go into detail on methods used to to make the computation of sparse loadings faster. We conclude Section \ref{sec2} with information on an extension for AR(1) errors in Section \ref{sec2.5} and how we tune for the number of factors to use in general and the $\ell_1$-norm penalty parameter of EM-sparse in Section \ref{sec2.6}. In Section \ref{sec3} we outline the structure of the package and functions it includes. Section \ref{sec4} provides two real-data case studies on UK inflation and trade in goods to display the package capabilities and visualisation features. Section \ref{sec5} concludes the article.

\section{Estimation Methods and Tuning}\label{sec2}
For a $p$-dimensional collection of stationary time series variables $\bX_t$ over $t=1,\dots,n$, we can assume the dynamic factor model for $r$ factors $\bF_t$ to be given by 
\begin{align}
    \bX_t &= \blam \bF_t + \beps_t \, , \label{eq:observation}\\
    \bF_t &= \bA\bF_{t-1} + \bu_t \, . \label{eq:factor}
\end{align}
The key assumption for this to be a sparse DFM is that the $p \times r$ loading parameter $\blam$ is sparse. The idiosyncratic errors $\{\beps_t\}$ are assumed to be multivariate white noise and for simplicity we assume $\E[\beps_t\beps_t^\top] = \bsig_{\beps} = \text{diag}(\bm{\sigma}^2_{\beps})$ and $\bm{\sigma}^2_{\beps} \in \mathbb{R}^p_+$ is a vector of idiosyncratic variances. In Section \ref{sec2.5} we give an extension to AR(1) idiosyncratic errors. We let $\E[\bu_t\bu_t^\top] = \bsig_{\bu}$ and assume $\|\bA\|_2 <1$, thus the latent VAR(1) model is stationary. This model corresponds to an exact DFM, where all the temporal dependence is modelled via the latent factors.

We now provide the four estimation algorithms the \textbf{sparseDFM} package can implement to obtain estimates of latent factors and model parameters. In the main \verb+sparseDFM()+ function the choice of estimation algorithm can be set using either \verb+alg = "PCA"+, \verb+alg = "2Stage"+, \verb+alg = "EM"+ or \verb+alg = "EM-sparse"+.

\subsection[PCA - Stock and Watson (2002)]{PCA - \cite{stock2002forecasting}} \label{sec2.1}
The classic way to estimate latent factors is through principal components analysis (PCA), which has been extensively studied in the literature \citep{stock2002forecasting, bai2003inferential, doz2020dynamic}. For centered and complete\footnote{$\bX_t$ can be made complete by simply removing rows with missing observations or by interpolation. In the package, we interpolate internal missing data using cubic splines and fill in missing data at the start and end of the sample using the median of the series which is then smoothed with an MA(3) process.} $\bX_t$ and assuming the number of $r$ factors is known\footnote{\cite{bai2002determining}, \cite{onatski2009testing} and \cite{ahn2013eigenvalue}, among others, have developed various approaches to
consistently estimate the number of factors. In the \textbf{sparseDFM} package we favour the information criteria approach of \cite{bai2002determining}. For more discussion see Section \ref{sec2.5}.}, PCA allows us to simultaneously estimate the factors and their loadings by solving the least squares problem:
\begin{equation}
    \min_{\blam,\bF} \frac{1}{np}\sum_{t=1}^n (\bX_t - \blam \bF_t)^\top (\bX_t - \blam \bF_t) = \min_{\blam,\bF} \frac{1}{np}\sum_{t=1}^n \beps_t^\top \beps_t \, ,
    \label{pca}
\end{equation}
subject to the normalization $\blam^\top \blam/p = \bI_r$ and $\bsig_F$ is diagonal (where $\bsig_F = \E(\bF_t\bF_t^\top)$). The solution to this least squares problem is the PC estimator of the factors $\hat{\bF}_t = p^{-1}\hat{\blam}^\top \bX_t$ where the estimated loadings $\hat{\blam}$ are the matrix of eigenvectors of the sample covariance matrix of $\bX_t$ associated with the largest $r$ eigenvalues. 

When mild conditions are placed on the correlation structure of idiosyncratic disturbances, the PCA estimator is the optimal non-parametric estimator for a large approximate DFM. With even tighter conditions of spherical idiosyncratic components (i.i.d. Gaussian), the PCA estimator is equivalent to the maximum likelihood estimator \citep{doz2020dynamic}. The problem with using non-parametric PCA methods to estimate DFMs is that there is no consideration of the dynamics of the factors or idiosyncratic components. In particular, there is no feedback from the estimation of the factor equation (\ref{eq:factor}) to the observable equation (\ref{eq:observation}). For this reason, it is preferable to use parametric methods that are able to model the dynamics in the system.

\subsection[2Stage - Giannone et al. (2008)]{2Stage - \cite{giannone2008nowcasting}} \label{sec2.2}
\cite{giannone2008nowcasting} proposed a two-stage (2Stage) framework for estimating DFMs which has since been theoretically studied by \cite{doz2011two} and successfully applied to the field of nowcasting. This method involves the following two steps:
\begin{itemize}
    \item[Step 1:] Preliminary estimates of the loadings $\blam$ and the factors $\bF_t$ are found via PCA as in (\ref{pca}). $\bsig_e$ is estimated as the empirical covariance of $\hat{\beps}_t = \bX_t - \hat{\blam}^\top \hat{\bF}_t$. The remaining parameters are estimated by fitting a VAR(1) model on $\hat{\bF}_t$.
    \item[Step 2:] The model is cast into state space form (\ref{eq:observation})-(\ref{eq:factor}) and the factors are re-estimated using the Kalman filter and smoother \citep{shumway1982approach, koopman2000fast}. 
\end{itemize}

The Kalman filter (KF) enables the factor estimates to be continually updated as new observations become available. This way, we are able to assess the impact new data releases of certain variables have in forecasting models. In the nowcasting literature this is referred to as the `news' of data releases. The KF provides the basis for smoothing, where the factor at a time $t$ can be estimated based on all data in the sample. This is very helpful for missing data interpolation and backcasting.

Appendix \ref{appendixA} provides the classic multivariate Kalman filter and smoother (KFS) equations of \cite{shumway1982approach}, often used in DFM estimation. We also provide the univariate treatment (sequential processing) of the multivariate equations for fast Kalman filter and smoother seen in \cite{koopman2000fast} that can lead to substantial computational gains for exact DFMs. As it is assumed in exact DFMs that $\bsig_{\beps}$ is diagonal, it becomes possible to filter the observations $\bX_t$ one element at a time, as opposed to altogether as in the classic multivariate approach. This way, matrix inversions become scalar divisions and thus huge speedups are possible. In the main \verb+sparseDFM()+ function it is possible to set \verb+kalman = "multivariate"+ or \verb+kalman = "univariate"+ for the respective KFS methods. The default is \verb+kalman = "univariate"+ for fast solutions when the errors are assumed i.i.d..

\subsection[EM - Banbura and Modugno (2014)]{EM - \cite{banbura2014maximum}} \label{sec2.3}
\cite{banbura2014maximum} build on the DFM representation of \cite{watson1983alternative} and adopt an expectation-maximisation (EM) approach to estimate the system (\ref{eq:observation})-(\ref{eq:factor}) by quasi-maximum likelihood estimation (QMLE)\footnote{It is described as quasi-maximum likelihood as consistency results are robust to cross-sectional misspecification, serial correlation of idiosyncratic components, and non-Gaussianity.}. Assuming residuals in the system (\ref{eq:observation})-(\ref{eq:factor}) are Gaussian, and collecting all parameters in $\btheta = (\blam, \bA, \bsig_{\beps}, \bsig_{\bu})$, the idea is to write the joint log-likelihood of $\bX_t$ and $\bF_t$ for all $t=1,\dots,n$ in terms of both the observed and unobserved data as:
\begin{align}
\label{eq:jointlogl}
     \log\mathcal{L}(\bX,\bF;\btheta)
      = &-\frac{1}{2}\log \vert\mathcal{\bP}_0\vert - \frac{1}{2}(\bF_0 - \bm{\alpha}_0)^\top \mathcal{\bP}_0^{-1} (\bF_0 - \bm{\alpha}_0) \nonumber \\
     & - \frac{n}{2} \log \vert\bsig_{\bu}\vert - \frac{1}{2}\sum_{t=1}^n \bu_t^\top \bsig_{\bu}^{-1} \bu_t \nonumber \\
     & - \frac{n}{2} \log \vert\bsig_{\beps}\vert - \frac{1}{2}\sum_{t=1}^n \beps_t^\top \bsig_{\beps}^{-1} \beps_t \, ,
\end{align}
where $\beps_t = \bX_t-\blam \bF_t$, $\bu_t=\bF_t-\bA\bF_{t-1}$, and we have assumed an initial distribution at $t=0$ of the factors as $\bF_0 \sim N(\bm{\alpha}_0, \mathcal{\bP}_0)$. We then iterate between the two steps:
\begin{itemize}
    \item[E Step:] Compute the expectation of the joint log-likelihood (\ref{eq:jointlogl}) conditional on the available information up to $n$, which we denote by $\bomega_n$, using the parameters estimated in the previous iteration, $j$:
    \begin{equation*}
        \E\left[ \log\mathcal{L}(\bX,\bF;\btheta^{(j)})|\bomega_n \right] \, ;
    \end{equation*}
    \item[M Step:] Re-estimate the parameters through the maximisation of the conditional expected joint log-likelihood:
    \begin{equation*}
        \btheta^{(j+1)} = \argmax_{\btheta} \E\left[ \log\mathcal{L}(\bX,\bF;\btheta^{(j)})|\bomega_n \right] \, ,
    \end{equation*}
\end{itemize}
until convergence\footnote{We choose to use the same EM convergence criteria as seen in \cite{doz2012quasi}, which is determined when:
\begin{equation}
\label{eq:loglconverge}
    M_j = \frac{\log\mathcal{L}(\bX;\hat{\btheta}^{(j)})-\log\mathcal{L}(\bX;\hat{\btheta}^{(j-1)})}{\left( \log\mathcal{L}(\bX;\hat{\btheta}^{(j)}) + \log\mathcal{L}(\bX;\hat{\btheta}^{(j-1)}) \right)\Big/2} < \text{threshold} \, ,
\end{equation}
where the user has on option in the sparseDFM() function to set the threshold value. The default is $10^{-4}$.}.

The first iteration of the EM algorithm requires initial parameter values $\btheta^{(0)}$. These are obtained by doing stage one of the 2Stage framework of \cite{giannone2008nowcasting} as in Section \ref{sec2.2}. The initial factor mean $\bm{\alpha}_0$ is set to 0 as in \cite{banbura2014maximum}. The initial factor covariance $\mathcal{\bP}_0$ is set to the matrix of $\mathrm{vec}(\mathcal{\bP}_0) = (\bI_{rp} - \hat{\bA} \otimes \hat{\bA})^{-1}\mathrm{vec}(\hat{\bsig}_{\bu})$. This initialisation will likely to be good when $p$ is large, as PCA estimation of factor loadings have been shown to be consistent for large cross-sections \citep{bai2003inferential, doz2011two}.

The EM algorithm then proceeds iterating between the E step and M step. Assuming Gaussian errors, the conditional expectation in the E step can be found using the Kalman filter and smoother equations of the state-space (\ref{eq:observation})-(\ref{eq:factor}) as given in Appendix \ref{appendixA}. From this we obtain the conditional mean and covariances of the state which we denote by:
\begin{align*}
    \ba_{t|n} &= \E[\bF_t|\bomega_n] \, ,\\
    \bP_{t|n} &= \mathrm{Var}[\bF_t|\bomega_n] \, ,\\
    \bP_{t,t-1|n} &= \mathrm{Cov}[\bF_t, \bF_{t-1}|\bomega_n] \, ,
\end{align*}
conditional on all information we have observed up to $n$, denoted by $\bomega_n$.

As shown in \cite{banbura2014maximum} and fully derived in Appendix \ref{appendixB}, the maximisation of the conditional expected log-likelihood (i.e. the M step) results in the following expressions for the parameter estimates:
\begin{equation*}
    \hat{\bm{\alpha}}_0 = \ba_{0\vert n}\quad;\quad \hat{\mathcal{\bP}}_0 =\bP_{0\vert n}    \label{eq:init} 
\end{equation*}
and letting $\bS_{t\vert n} = \ba_{t\vert n}\ba_{t\vert n}^{\top} + \bP_{t\vert n}$, and $\bS_{t,t-1\vert n} = \ba_{t\vert n}\ba_{t-1\vert n}^{\top} + \bP_{t,t-1\vert n}$ we have
\begin{align*}
    \hat{\bA} &= \left(\sum_{t=1}^n \bS_{t-1\vert n}\right)^{-1} \left(\sum_{t=1}^n \bS_{t,t-1\vert n}\right)  \, , \\
    \hat{\bsig}_{\bu} &= \frac{1}{n} \sum_{t=1}^n\left[\bS_{t\vert n} - \hat{\bA} \left( \bS_{t-1,t\vert n}  \right) \right] \, . \label{sigu}
\end{align*}

For parameters $\blam$ and $\bsig_{\beps}$, we should also consider there might be missing data in $\bX_t$. \cite{banbura2014maximum} define a selection matrix $\bW_t$ to be a diagonal matrix such that 
\begin{equation*}
W_{t,ii}=\begin{cases}
1 & \mathrm{\mathrm{if}\;}X_{i,t}\;\mathrm{observed}\\
0 & \mathrm{\mathrm{if}\;}X_{i,t}\;\mathrm{missing}
\end{cases}
\end{equation*}
and note that $\bX_t = \bW_t\bX_t + (\bI-\bW_t)\bX_t $. The update for the loadings and the idiosyncratic error covariance is then given by
\begin{align*}
    \mathrm{vec}(\hat{\blam}) &= \left( \sum_{t=1}^n \bS_{t\vert n} \otimes \bW_t\right)^{-1} \mathrm{vec}\left( \sum_{t=1}^n \bW_t\bX_t\ba_{t|n}^\top\right) \, , \\[1em] 
    \hat{\bsig}_{\beps} &= \frac{1}{n}\sum_{t=1}^n \mathrm{diag}\Bigg[  \bW_t \bigg( \bX_t\bX_t^\top\ - 2\bX_t \ba_{t\vert n}^\top \hat{\blam}^\top 
     + \hat{\blam}\bS_{t\vert n}\hat{\blam}^\top\bigg)\bW_t + (\bI - \bW_t)\hat{\bsig}_{\beps}^*(\bI - \bW_t) \Bigg] \, , \label{eq:sigeps}
\end{align*}
where $\mathrm{vec}(\cdot)$ represents vectorisation of a matrix and $\hat{\bsig}_{\beps}^*$ is obtained from the previous EM iteration. 
In practice we update $\hat{\bsig}_{\beps}$ after estimating $\hat{\blam}$, as the former is based on the difference between the observations and the estimated common component. See Appendix \ref{appendixB} for full detail of the EM algorithm. 

\cite{doz2012quasi} and \cite{barigozzi2019quasi} provide consistency results of the EM approach as $(n,p) \rightarrow \infty$. 
In terms of macroeconomic forecasting, the EM algorithm approach has many
gains over the PCA and 2Stage procedures. It is much more efficient in small samples,
as shown by simulation studies of \cite{doz2012quasi}. It is a very flexible framework able
to deal with arbitrary patterns of missing data in $\bX_t$, as shown in \cite{banbura2014maximum}. It enables one to impose restrictions on parameters when updating them in the EM algorithm. For example, \cite{banbura2010nowcasting} impose restrictions on the loadings to reflect temporal aggregation
and introduce factor specific groups of variables in their nowcasting application. Furthermore, it is a computationally efficient algorithm allowing fast forecasts for policy
makers/analysts, as opposed to computationally demanding MCMC based alternatives. 

\subsection[EM-sparse - Mosley et al. (2023)]{EM-sparse - \cite{mosleySAC}} \label{sec2.4}
\cite{mosleySAC} extend the \cite{banbura2014maximum} EM algorithm framework by imposing LASSO regularisation on the conditional expected log-likelihood for the $\blam$ update with the endeavour of obtaining sparse factor loadings. In particular, they find estimators of the form 
\begin{equation}
    \hat{\btheta} = \arg\min_{\btheta} \left[-\E \left[\log\mathcal{L}(\bX,\bF;\btheta)\vert\bomega_n\right] + \alpha \|\blam\|_1\right] \;, \label{eq:reg_elik}
\end{equation}
where $\alpha$ is the LASSO tuning parameter controlling the amount of sparsity in $\blam$. Due to the complexity of the objective being maximised in (\ref{eq:reg_elik}), \cite{mosleySAC} propose to solve the LASSO problem using the Alternative Directed Method of Moments (ADMM) by \cite{boyd2011distributed} as this approach provides a way to split up the complex objective function into separate, simpler optimisation problems. This involves considering the penalised and augmented Lagrangian 
\begin{equation}
    \mathcal{C}(\blam, \bZ, \bU) := -\E\left[\log\mathcal{L}(\bX,\bF;\btheta)|\bomega_n\right] +\alpha\|\bZ\|_{1}+\frac{\nu}{2}\|\blam-\bZ+\bU\|_{F}^{2} \, ,\label{eq:augmented}
\end{equation}
where $\bZ\in\mathbb{R}^{p\times r}$ is an auxiliary variable, $\bU\in\mathbb{R}^{p\times r}$
are the (scaled) Lagrange multipliers and $\nu$ is the scaling term. Under equality conditions relating the auxilary ($\bZ$) to the primal ($\blam$) variables, this is equivalent to minimising (\ref{eq:reg_elik}), e.g.
\begin{align*}
&\arg\min_{\bZ=\blam} \max_{\bU}\mathcal{C}(\blam, \bZ, \bU) \\
&= \arg\min_{\blam}\left[-\E \left[\log\mathcal{L}(\bX,\bF;\btheta)\vert\bomega_n\right] + \alpha \|\blam\|_1\right] \, ,
\end{align*}
as (\ref{eq:reg_elik}) is convex in the argument $\blam$ with all other parameters fixed, this argument holds for any $\nu>0$.

The augmented Lagrangian (\ref{eq:augmented}) can be sequentially minimised via the following updates\footnote{The full derivation of these equations are given in Appendix \ref{appendixC}. 
}
\begin{align*}
\blam^{(k+1)} & =\argmin_{\blam}\mathcal{C}(\blam,\bZ^{(k)},\bU^{(k)})\\
\bZ^{(k+1)} & =\argmin_{\bZ}\mathcal{C}(\blam^{(k+1)},\bZ,\bU^{(k)})\\
&=\mathrm{soft}(\blam^{(k+1)}+\bU^{(k)};\alpha/\nu)\\
\bU^{(k+1)} & =\bU^{(k)}+\blam^{(k+1)}-\bZ^{(k+1)} \, .
\end{align*}
for $k=0,1,2,\ldots,$ until convergence.
The first (primal) update is simply a least-squares type problem, whereby on vectorising $\blam$ one finds
\begin{equation}
\mathrm{vec}(\blam^{(k+1)}) = \left( \sum_{t=1}^n \bS_{t\vert n} \otimes \bW_t\bsig_{\beps}^{-1}\bW_t + \nu\bI_{pr} \right)^{-1}  
\mathrm{vec}\bigg[ \sum_{t=1}^n\bW_t\bsig_{\beps}^{-1}\bW_t\bX_t \ba_{t\vert n}^\top 
+  \nu(\bZ^{(k)}-\bU^{(k)})\bigg] \, .\label{eq:admmLam}
\end{equation}

\cite{mosleySAC} further detail how the computation of (\ref{eq:admmLam}) can be made significantly faster by exploiting dimensionality reduction, going from a per-iteration cost of the ADMM of $\mathcal{O}(r^3p^3)$ to  $\mathcal{O}(r^3p)$. For details of this speed up refer to \cite{mosleySAC}. We implement this speed up in the package. 

\subsection{An Extension for AR(1) Idiosyncratic Errors} \label{sec2.5}
To extend the state-space framework (\ref{eq:observation})-(\ref{eq:factor}) to AR(1) idiosyncratic errors, as opposed to i.i.d. white noise, we just follow the methodology laid out in \cite{banbura2014maximum} and augment the state vector so it contains the latent factors and the latent AR(1) errors.  Formally, we set up the AR(1) process like so 
\begin{align*}
    \beps_{i,t} &= \Tilde{\beps}_{i,t} + \bm{\eta}_{i,t} \, , \quad \bm{\eta}_{i,t} \sim N(0, \kappa) \, , \\
    \Tilde{\beps}_{i,t} &= \phi_i\Tilde{\beps}_{i,t-1} + \be_{i,t} \, , \quad \be_{i,t} \sim N(0, \sigma_i^2) \, ,
\end{align*}
for $i=1,\dots,p$ and $t=1,\dots,n$,  where $\kappa$ is a very small number which allows us to easily set up the new state-space framework. 

This new state-space framework is given by 
\begin{align*}
    \bX_t &= \Tilde{\blam}\Tilde{\bF}_t + \bm{\eta}_t \, , \quad \bm{\eta}_t \sim N(0, \bsig_{\bm{\eta}}) \, , \\
    \Tilde{\bF}_t &= \Tilde{\bA}\Tilde{\bF}_{t-1} + \Tilde{\bu}_t \, , \quad \Tilde{\bu}_t \sim N(0, \bsig_{\Tilde{\bu}}) \, , 
\end{align*}
where 
$$
\tilde{\bF}_t=\left[\begin{array}{c}
\bF_t \\
\tilde{\beps}_t
\end{array}\right] \, , \; \tilde{\bu}_t=\left[\begin{array}{l}
\bu_t \\
\be_t
\end{array}\right] \, , \; \tilde{\blam}=\left[\begin{array}{cc}
\blam & \bI
\end{array}\right] \, , \; \tilde{\bA}=\left[\begin{array}{cc}
\bA & \bm{0} \\
\bm{0} & \bphi
\end{array}\right] \, , 
$$
$$
\bsig_{\tilde{\bu}}=\left[\begin{array}{cc}
\bsig_{\bu} & \bm{0} \\
\bm{0} & \bsig_{\be}
\end{array}\right] \; \text{and} \quad  \bsig_{\bm{\eta}} = \kappa \bI_{r+p} \, , 
$$
for $\bphi = \text{diag}(\phi_1, \dots, \phi_p)$ and $\bsig_{\be} = \E[\be_t\be_t^\top] = \text{diag}(\bm{\sigma}_{\be}^2)$ and $\bm{\sigma}_{\be}^2 \in \mathbb{R}^p_{+}$ is a vector of error variances. 

The parameter estimates of this new augmented state-space model can be obtained very similarly to before. See \cite{banbura2014maximum} for the estimates of the new set of parameters $\tilde{\btheta}=(\bA, \bphi, \blam, \bsig_{\bu}, \bsig_{\be})$.

\subsection{Tuning the Model} \label{sec2.6}
\subsubsection{For the number of factors}
To calculate the number of factors to use in the model we opt to take the information criteria approach of \cite{bai2002determining}. This can be done before the \verb+SparseDFM()+ model is fitted. \cite{bai2002determining} consider 3 information criteria with different penalties of the form:
\begin{align*}
    IC_1(r) &= \log\left(V_r(\hat{\bF},\hat{\blam})\right) + r \left( \frac{n+p}{np}\right)\log\left( \frac{np}{n+p}\right) \, , \\
    IC_2(r) &= \log\left(V_r(\hat{\bF},\hat{\blam})\right) + r \left( \frac{n+p}{np} \right)\log\left( \min\{n,p\}\right) \, , \\
    IC_3(r) &= \log\left(V_r(\hat{\bF},\hat{\blam})\right) + r \frac{\log\left( \min\{n,p\}\right)}{\min\{n,p\}} \, .
\end{align*}
The sum of squared residuals for $r$ factors $V_r(\hat{\bF},\hat{\blam}) = \sum_{i=1}^p\sum_{t=1}^n\E[\hat{\epsilon}_{i,t}^2]\big/{np}$ with $\hat{\epsilon}_{i,t} = X_{t,i}-\hat{\bF}_t\hat{\blam}_i$ is found using PCA on the standardised dataset $\bX$. The estimated factors $\hat{\bF}$ corresponding to the principal components and the estimated loadings $\hat{\blam}$ corresponding to the eigenvectors. Should the data contain missing values, we first interpolate the missing values using the median of the series that are then smoothed by a simple moving window as discussed in Section \ref{sec2.3}.

The number of factors to use will correspond to $\argmin_r IC_i(r)$ for $i=1,2$ or $3$. The second penalty is the highest when working in finite samples and therefore is set to default in our \verb+tuneFactors()+ function. 

\subsubsection{For the LASSO sparsity parameter}
We tune the model for the best $\alpha$ by doing a grid search over a range of $\alpha$'s using BIC. For factors $\hat{\bF}$ and $\hat{\blam}$ estimated by a run of the EM algorithm for a particular $\alpha$ value, $\alpha^*$, the BIC is 
\begin{equation*}
    BIC_{\alpha^*} = \log\left( V_{\alpha^*}(\hat{\bF},\hat{\blam}) \right) + m \left(\frac{\log(np)}{np} \right) \, ,
\end{equation*}
where $V_{\alpha^*}(\hat{\bF},\hat{\blam})$ is the sum of squared residuals using $\alpha^*$ and $m$ is the number of non-zero entries in $\hat{\blam}$. 

The default search grid for $\alpha$ is a log-space from $10^{-2}$ to $10^3$. Any grid can be inputted by the user and we make sure the grid is ordered from low to high. Each time the EM algorithm is run with a new $\alpha$ value, the estimated parameters from the previous $\alpha$ value are used as a \emph{warm start} to the EM algorithm. This should reduce the number of iterations needed for the EM algorithm to converge as we search through the $\alpha$ grid. We implement a stopping rule in the grid search if a column of the loadings parameter $\blam$ gets set entirely to zero.

\section[Package Structure]{Package Structure}\label{sec3}
\subsection{Installation}
The \textbf{sparseDFM} package can be installed from CRAN as follows:
\begin{verbatim}
    install.packages("sparseDFM")
\end{verbatim}
The development version can be installed from GitHub using the \textbf{devtools} package \citep{devtools}:
\begin{verbatim}
    devtools::install_github("mosleyl/sparseDFM")
\end{verbatim}
To load the package into the R environment, type:
\begin{verbatim}
    library(sparseDFM)
\end{verbatim}

\subsection{Core Function}
The core function of the package is \verb+sparseDFM+. This function allows different estimation methods for dynamic factor models including: principal components analysis (PCA) for static factors \citep{stock2002forecasting},  the two-stage (2Stage) framework of PCA plus Kalman filter/smoother \citep{giannone2008nowcasting},  the quasi-maximum likelihood approach using the expectation-maximisation (EM) algorithm to handle arbitrary patterns of missing data \citep{banbura2014maximum}, and the novel sparsified EM approach (EM-sparse) allowing LASSO regularisation on factor loadings \citep{mosleySAC}. The function can be implemented using:
\begin{verbatim}
    sparseDFM(X, r, q = 0, alphas = logspace(-2,3,100), alg = "EM-sparse", 
              err = "IID",  kalman = "univariate", store.parameters = FALSE,
              standardize = TRUE, max_iter = 100, threshold = 1e-4)
              
\end{verbatim}
It takes the following arguments:
\begin{itemize}
    \item \verb+X+: An $n \times p$ numeric data matrix or data frame representing $p$ stationary time series variables with $n$ observations. Missing data should be represented by \verb+NA+. 
    \item \verb+r+: The number of factors the model should estimate. This can be selected using the function \verb+tuneFactors()+. Usually less than 10 factors should be sufficient. 
    \item \verb+q+: The first \verb+q+ variables in \verb+X+ will not be regularised. This parameter is only relevant when \verb+alg = "EM-sparse"+. It is useful when you want to ensure variables are always loaded onto all factors and hence regularisation is not applied to these variables. Set the first \verb+q+ series in \verb+X+ to be these variables. The default is \verb+q = 0+ where all variables are regularised. 
    \item \verb+alphas+: Numeric vector of $\ell_1$-norm penalty parameters. This parameter is only relevant when \verb+alg = "EM-sparse"+. The default grid is set to be \verb+alphas = logspace(-2, 3, 100)+ to ensure a wide range of $\alpha$ values. Each iteration of the EM algorithm uses the previous $\alpha$'s parameter estimates as a warm-start to the EM algorithm. The grid search will stop when all of the values in \verb+alphas+ are used for estimation or when a column of the estimated loadings $\hat{\blam}$ becomes entirely 0. This will occur if $\alpha$ is too high. Single $\ell_1$-norm penalty parameters can be used as well as vectors. 
    \item \verb+alg+: The choice of estimation algorithm. Options include:
    \begin{itemize}
        \item[-] \verb+"PCA"+: principal components analysis (PCA) for static factors \citep{stock2002forecasting};
        \item[-] \verb+"2Stage"+: the two-stage framework of PCA plus Kalman filter/smoother \citep{giannone2008nowcasting};
        \item[-] \verb+"EM"+: The quasi-maximum likelihood approach using the EM algorithm to handle arbitrary patterns of missing data \citep{banbura2014maximum};
        \item[-] \verb+"EM-sparse"+: The novel sparsified EM approach allowing LASSO regularisation on factor loadings \citep{mosleySAC}. This is the default. 
    \end{itemize}
    \item \verb+err+: The choice of distribution for idiosyncratic residuals. Options are \verb+err = "IID"+ for IID white noise residuals and this is the default. Or \verb+err = "AR1"+ for AR(1) residuals. It is recommended to use IID errors for computational efficiency. See Appendix \ref{appendixH} for details.  
    \item \verb+kalman+: The choice of Kalman filter and smoother equations to use. Options include:
    \begin{itemize}
        \item[-] \verb+"multivariate"+: classic Kalman filter and smoother equations seen in  \cite{shumway1982approach}.
        \item[-] \verb+"univariate"+: univariate treatment (sequential processing) of the multivariate equations for fast Kalman filter and smoother seen in \cite{koopman2000fast}. This is the default.
    \end{itemize}
    It is recommended to use \verb+kalman = "univariate"+ when \verb+err = "IID"+ and \verb+kalman = "multivariate+ when \verb+err = "AR1"+. See Appendix \ref{appendixH} for details. 
    \item \verb+store.parameters+: A logical parameter indicating whether the estimation output of parameters and factors should be stored for every value of $\alpha$ considered in \verb+alphas+. This parameter is only relevant when \verb+alg = "EM-sparse"+. This may be useful when checking alternative outputs other than the one for the optimal $\alpha$. The default is \verb+store.parameters = FALSE+. 
    \item \verb+standardize+: A logical parameter indicating whether the data should be standardised or not before estimating the model. The default is \verb+standardized = TRUE+.
    \item \verb+max_iter+: An integer representing the maximum number of EM iterations. The default is \verb+max_iter = 100+. 
    \item \verb+threshold+: Tolerance on EM iterates. See equation (\ref{eq:loglconverge}). Default is \verb+threshold = 1e-4+.
\end{itemize}

The function \verb+sparseDFM+ returns an S3 object of class \verb+`sparseDFM'+. This allows the user to pass the generic functions: \verb+print()+, \verb+summary()+, \verb+plot()+, \verb+residuals()+, \verb+resids()+, \verb+fitted()+, and \verb+predict()+ to the returned \verb+sparseDFM+ object to display generic outputs. The \verb+predict()+ function will itself return an S3 object of class \verb+`sparseDFM_forecast'+ to allow the user to \verb+print()+ forecast results. There are plenty of helpful types of plot included in the package to allow the user to visualise estimation results (see Appendix \ref{appendixD}). We go into detail on the specific usage of these options in the empirical applications of Section \ref{sec4}.  

When called, the S3 object \verb+sparseDFM+ returns a list-of-lists containing the following:
\begin{itemize}
    \item \verb+data+: A list containing information on the data used for the model fit. Including the original data matrix, the initial balanced data matrix from \verb+fillNA()+, the fitted values $\hat{\blam}\hat{\bF}$ and more. 
    \item \verb+params+: A list containing the estimated model parameters of $\hat{\bA}$, $\hat{\blam}$, $\hat{\bsig}_u$ and $\hat{\bsig}_\epsilon$. 
    \item \verb+state+: A list containing the estimated states and state covariances for each time point considered. I.e. the factors and factor covariances, along with AR(1) errors and covarainces if \verb+err = AR1+. 
    \item \verb+em+: A list containing information about the convergence of the EM algorithm and $\ell_1$-norm penalty parameter tuning. 
    \item \verb+alpha.output+: Parameter and state outputs for each $\ell_1$-norm penalty parameter in \verb+alphas+. Only stored if \verb+store.parameters = TRUE+.
\end{itemize}

\subsection{Other Functions}
This package also includes eight other exported functions:
\begin{center}
    \begin{tabular}{|p{4cm}|p{11cm}|}
\hline 
    \textbf{Function} & \textbf{Brief Description} \\ \hline 
    \verb+kalmanMultivariate()+ & Function to run the classic multivariate Kalman filter and smoother equations of \cite{shumway1982approach}. \\ \hline 
    \verb+kalmanUnivariate()+ & Function to run the fast univariate Kalman filter and smoother equations of \cite{koopman2000fast}. \\ \hline  
    \verb+fillNA()+ & Function to interpolate missing data in $\bX$. Used to obtain a balanced dataset for the PCA initialisation fit.  \\ \hline 
    \verb+transformData()+ & Function to transform a non-stationary data matrix to become stationary by taking first/second (log) differences/growth rates. \\ \hline 
    \verb+tuneFactors()+ & Function to determine the number of factors to use based on the \cite{bai2002determining} information criteria and variance explained from PCA. \\ \hline 
    \verb+missing_data_plot()+ & Visualise the amount of missing data in a data matrix or data frame. \\ \hline 
    \verb+raggedEdge()+ & Function to generate a ragged edge structure for the end of sample of a data matrix/frame. \\ \hline 
    \verb+logspace()+ & Function to produce a vector of log10 space values. \\ \hline
\end{tabular}
\end{center}

\section[Using the sparseDFM Package]{Using the sparseDFM Package}\label{sec4}

In this section we provide a tutorial of how one might use the package for empirical applications to real data. We provide two case studies both with the intention to display the package capabilities and plotting features. The first case study focusses on a small subset of \textbf{quarterly CPI} (consumer price inflation) index data for the UK taken from the Office from National Statistics' (ONS) website\footnote{We use the Q4 2022 release (benchmarked to 2015=100) from \url{https://www.ons.gov.uk/economy/inflationandpriceindices/datasets/consumerpriceindices}}. The data contains 36 variables of different classes of the inflation index and 135 observations from 1989 Q1 to 2022 Q3. The purpose of this small, 36 variable example is to demonstrate the core functionality and the ability to graphically display results with the \textbf{sparseDFM} package. The second case study applies the package onto a large-scale data set for the purpose of \textbf{nowcasting UK trade in goods}. The data contains 445 columns, including 9 target series (UK exports of the 9 main commodities worldwide) and 434 monthly indicator series, and 226 rows representing monthly values from January 2004 to October 2022. In this large scale example we make comparisons between a regular DFM and a Sparse DFM in terms of interpreting factor structure and accuracy of predictions. 

Before going through the case studies, we refer the reader to the table in Appendix \ref{appendixD} that summarises the various types of plot and other S3 generic functions that the package contains.

\subsection{Case Study 1: Exploring UK Inflation}
The inflation data is included in the \textbf{sparseDFM} package and we can load it into the R environment as a data frame called \verb+data+ using:
\begin{verbatim}
  R> data <- inflation
\end{verbatim}
Before we fit a model it is worthwhile to first perform some exploratory data analysis. Two main properties to look out for when working with dynamic factor models are the amount of missing data present and if the data series are stationary or not. The function \verb+missing_data_plot()+ allows the user to visualise where missing data is present:
\begin{verbatim}
    R> missing_data_plot(data)
\end{verbatim}
\begin{figure}[h!]
    \centering
    \includegraphics[width=0.8\linewidth]{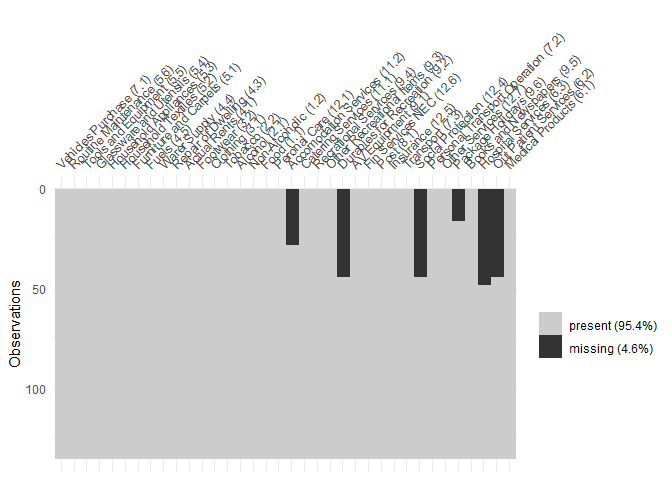}
    \caption{Output of \emph{missing\_data\_plot(data)} showing the amount of missing data in the inflation example.}
    \label{fig:missdata}
\end{figure}
Figure \ref{fig:missdata} shows the output of this missing data plot. We can see that $6$ of the $36$ variables have missing data towards the start of the sample.

In a DFM, we assume the latent factors to follow a stationary VAR(1) process and hence we require the input data to be stationary. After exploration it appears the inflation  data is stationary in first-differences. Hence, we can apply the function \verb+transformData()+ and set the \verb+stationary_transform+ parameter to represent a first-difference transform for all variables:
\begin{verbatim}
    R> new_data <- transformData(data, stationary_transform = rep(2, ncol(data))))
\end{verbatim}
We store the stationary first-difference data as the variable \verb+new_data+. The \verb+stationary_transform+ parameter is a vector of length matching the number of variables in the data that contains a stationary transform index for each variable with the options: \verb+1+ (no change), \verb+2+ (first-difference), \verb+3+ (second-difference), \verb+4+ (log-first-difference), \verb+5+ (log-second-difference), \verb+6+ (growth rate) or \verb+7+ (log growth rate). The returned transformed data set is of the same dimension as the original data with \verb+NA+ where missing data is, for example, the first row of \verb+new_data+ will all be \verb+NA+ as we do not have an observed 1988 Q4 value to difference with. 

Now we have a stationary data set, the next step would be to determine the number of factors to use with our DFM. To do this we can apply the function \verb+tuneFactors()+ that selects the best number of factors to use based on \cite{bai2002determining} information criteria:
\begin{verbatim}
    R> tuneFactors(new_data, type = 2, standardize = TRUE, 
                  r.max = min(15,ncol(X)-1), plot = TRUE)
\end{verbatim}
 The default information criteria (IC) is \verb+type = 2+ (see Section \ref{sec2.6} for details). We search for a maximum number of 15 factors and plot the results. Figure \ref{fig:tunefactors} (in Appendix \ref{appendixE}) shows the output of \verb+tuneFactors()+: the IC values for different number of factors along with a screeplot showing the percentage variance explained obtained from the eigenvalues of the covariance of the data. 

\begin{figure}[h!]
    \centering
    \includegraphics[width=0.8\linewidth]{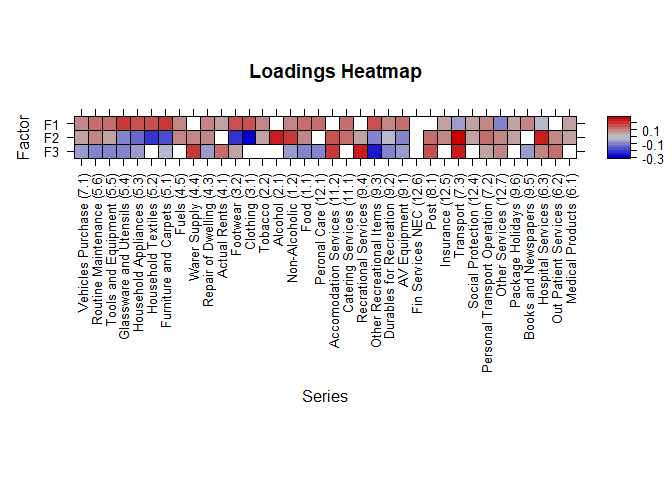}
    \caption{Output of \emph{plot()} with \emph{type = ``loading.heatmap"} showing a heatmap of the estimated loadings from Sparse DFM fit on the inflation data.}
    \label{fig:loadingplot}
\end{figure}

It appears 3 factors is sufficient for the inflation exercise. We can now fit a Sparse DFM using 3 factors and call the generic S3 function \verb+summary()+ to summarise the call:
\begin{verbatim}
fit.sdfm <- sparseDFM(new_data, r = 3, alphas = logspace(-2,3,100),
                      alg = 'EM-sparse', err = 'IID', kalman = 'univariate')
summary(fit.sdfm)
#> 
#> Call: 
#> 
#>  sparseDFM(X = new_data, r = 3, alphas = logspace(-2, 3, 100), 
              alg = "EM-sparse", err = "IID", kalman = "univariate")
#> 
#>  Sparse Dynamic Factor Model using EM-sparse with: 
#> 
#>  n = 135 observations, 
#>  p = 36 variables, 
#>  r = 3 factors, 
#>  err = IID
#> 
#>  The r x r factor transition matrix A 
#>            F1         F2          F3
#> F1  0.4654857  0.9849589 -0.96021652
#> F2  0.6433499 -0.3851340 -0.55318409
#> F3 -0.2577013  0.9030837 -0.02329362
#> 
#> 
#>  The r x r factor transition error covariance matrix Sigma_u 
#>            u1         u2         u3
#> u1  3.5334868 -0.2695235 -0.5345583
#> u2 -0.2695235  2.4666091  0.7245212
#> u3 -0.5345583  0.7245212  2.5614458
\end{verbatim}
We choose to tune the $\ell_1$-norm penalty parameter using BIC over the default logspace grid of $100$ values between $10^{-2}$ and $10^3$. The algorithm will stop searching over this grid as soon as an entire column of $\blam$ becomes $0$. We can find the optimal $\ell_1$-norm parameter $\hat{\alpha}$ and plot the BIC curve (shown in Figure \ref{fig:biccurve} in Appendix \ref{appendixE}) using:
\begin{verbatim}
# The best alpha chosen 
fit.sdfm$em$alpha_opt
#> [1] 0.4641589
# Plot the BIC values for each alpha 
plot(fit.sdfm, type = 'lasso.bic')
\end{verbatim}

We are able to display the estimation results using various types of plots - see Appendix \ref{appendixD} for options. We can plot the three estimated factors overlaid on top of the standardised inflation data  using 
\begin{verbatim}
# Plot all of the estimated factors 
plot(fit.sdfm, type = 'factor')
\end{verbatim}
and make a boxplot of the residuals for each inflation series using 
\begin{verbatim}
# Plot boxplots for the residuals of each variable 
plot(fit.sdfm, type = 'residual', use.series.names = TRUE)
\end{verbatim}
See Figure \ref{fig:factorest} and \ref{fig:residualest} in Appendix \ref{appendixE} for these plots, respectively. Note, we can set \verb+use.series.names = FALSE+ to avoid plotting the series names. A very useful plot to understand the factor loading structure is to use:
\begin{verbatim}
# Plot a heatmap of the estimated loadings 
plot(fit.sdfm, type = 'loading.heatmap', use.series.names = TRUE)
\end{verbatim}
This produces a heatmap (shown in Figure \ref{fig:loadingplot}) of the estimated factor loadings with a clear display of sparse loadings displayed by blank squares. 

\subsection{Case Study 2 - Nowcasting UK Trade in Goods (Exports)}

\textbf{Nowcasting}\footnote{For a detailed survey on the nowcasting literature see \cite{banbura2013now}} is a method used in econometrics that involves estimating the current state of the economy based on the most recent data available. It is an important tool because it allows policy makers and businesses to make more informed decisions in real-time, rather than relying on outdated information due to publication delays that may no longer be accurate. Trade in Goods (imports and exports) is currently published with a \textbf{2 month lag} by the UK's Office for National Statistics (ONS), which is quite a long time to wait for current assessments of trade, especially during times of economic uncertainty or instability. Nowcasting UK trade information has become particularly important in recent years due to the key events of the \textbf{Brexit referendum}, held in 2016, and the \textbf{coronavirus pandemic}, reaching UK shores early 2020. While the cause of these shocks are drastically different, both have imposed restrictions on trade in goods.

Using \textbf{sparseDFM}, we consider the task of understanding and nowcasting the movements of 9 monthly target series representing 9 of the main commodities the UK exports worldwide. These include: \textbf{Food \& live animals}, \textbf{Beverages and tobacco}, \textbf{Crude materials}, \textbf{Fuels}, \textbf{Animal and vegetable oils and fats}, \textbf{Chemicals}, \textbf{Material manufactures}, \textbf{Machinery and transport} and \textbf{Miscellaneous manufactures}.

To try and produce accurate nowcasts for the targets we use a large collection of 434 monthly indicator series including:
\begin{itemize}
    \item \textbf{Index of Production (IoP)} - Movements in the volume of production for the UK production industries - \emph{2 month lag} - \emph{89 series}.
    \item \textbf{Consumer Price Inflation (CPI)} - The rate at which the prices of goods and services bought by households rise or fall - \emph{1 month lag} - \emph{166 series}.
    \item \textbf{Producer Price Inflation (PPI)} - Changes in the prices of goods bought and sold by UK manufacturers - \emph{1 month lag} - \emph{153 series}.
    \item \textbf{Exchange rates} - Sterling exchange rates with 12 popular currencies - \emph{1 month lag} - \emph{12 series}.
    \item \textbf{Business confidence Index (BCI)} - Opinion surveys on developments in production, orders and stocks of finished goods in the industry sector - \emph{1 month lag} - \emph{1 series}.
    \item \textbf{Consumer Confidence Index (CCI)} - Opinion surveys on future developments of households’ consumption and saving - \emph{1 month lag} - \emph{1 series}.
    \item \textbf{Google Trends (GT)} - Popularity scores of 14 google search queries related to trade in goods - \emph{real-time} - \emph{14 series}.
\end{itemize}

\begin{figure}
\vspace{-2em}
    \centering
    \includegraphics[width=0.9\linewidth]{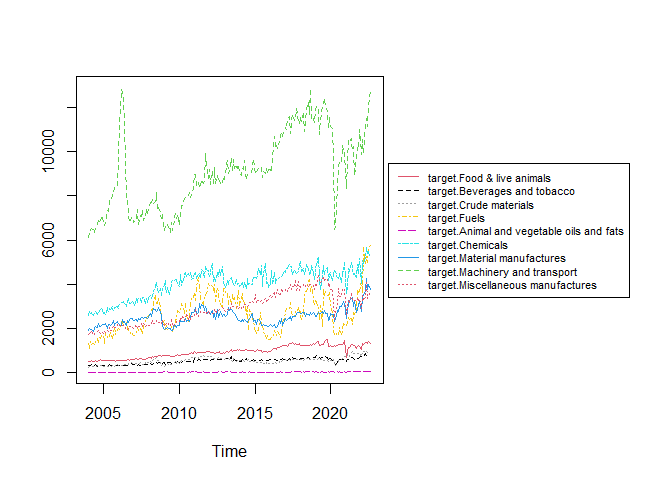}
    \caption{The target series of the 9 main commodities of UK exports worldwide.}
    \label{fig:exportseries}
\end{figure}

The \textbf{sparseDFM} package contains a data frame \verb+exports+ containing the 9 monthly target series of UK exports and the 434 monthly indicator series observed from January 2004 to October 2022\footnote{All data is openly available from the Office for National Statistics' website: \url{https://www.ons.gov.uk/}.} and can be loaded into the R environment using \begin{verbatim}
  R> data <- exports
\end{verbatim}
Figure \ref{fig:exportseries} provides a plot of the 9 target series we are interested in nowcasting. We see exports of machinery and transport being the largest. We also see two main drops during the 2009 and 2020 recessions and an upwards trend in the past year or so. This figure can be generated using:
\begin{verbatim}
# Plot the 9 target series using ts.plot with a legend on the right 
def.par <- par(no.readonly = TRUE) # initial graphic parameters 
goods <- data[,1:9]
layout(matrix(c(1,2),nrow=1), width=c(4,3)) 
par(mar=c(5,4,4,0)) 
ts.plot(goods, gpars= list(col=10:1,lty=1:10))
par(mar=c(5,0,4,2)) 
plot(c(0,1),type="n", axes=F, xlab="", ylab="")
legend("center", legend = colnames(goods), col = 10:1, lty = 1:10, cex = 0.7)
par(def.par) # reset graphic parameters to initial
\end{verbatim}

To make the data stationary, we apply first-difference transformations to each series using:
\begin{verbatim}
    R> new_data <- transformData(data, stationary_transform = rep(2, ncol(data))))
\end{verbatim}
The only missing data present in the data is at the end of the sample during the months of September and October 2022 depending on publication delays of the variables. This forms a ragged edge at the end of the sample with both months missing for the targets and IoP, just October missing for CPI, PPI, exchange rates, BCI and CCI, and nothing missing for google trends. This can be visualised with Figure \ref{fig:missdata12} in Appendix \ref{appendixF} which plots a \verb+missing_data_plot()+ for the last 12 months of the sample using:
\begin{verbatim}
# last 12 months 
data_last12 = tail(data, 12)
# Missing data plot. Too many variable names so use.names is set to FALSE for
# clearer output.
missing_data_plot(data_last12, use.names = FALSE)
\end{verbatim}

We are interested in nowcasting both September and October for the export targets. We tune for the number of factors using (see Figure \ref{fig:tunefactorsexports} in Appendix \ref{appendixF} for the output)
\begin{verbatim}
    R> tuneFactors(new_data, type = 2, standardize = TRUE, 
                  r.max = min(15,ncol(X)-1), plot = TRUE)
\end{verbatim}
According to the \cite{bai2002determining} information criteria, the best number of factors to use is 7. However, the screeplot seems to suggest that after 4 factors, the addition of more factors does not add that much in terms of explaining the variance of the data. For this reason, we choose to use 4 factors when modelling.

We now fit a regular DFM and a Sparse DFM to the data with 4 factors:
\begin{verbatim}
## Regular DFM fit - takes around 18 seconds 
fit.dfm <- sparseDFM(new_data, r = 4, alg = 'EM')

## Sparse DFM fit - takes around 2 mins to tune 
# set q = 9 as the first 9 variables (targets) should not be regularised
# L1 penalty grid set to logspace(0.4,1,15) after exploration
fit.sdfm <- sparseDFM(new_data, r = 4, q = 9, alg = 'EM-sparse',
                      alphas = logspace(0.4,1,15))
\end{verbatim}
Note, we set the parameter \verb+q = 9+ in the Sparse DFM fit as we do not want to regularise the first 9 variables of the data set corresponding to the 9 target series. 

We can explore the convergence and tuning of each algorithm like so:
\begin{verbatim}
# Number of iterations the DFM took to converge
fit.dfm$em$num_iter
#> [1] 14

# Number of iterations the Sparse DFM took to converge at each L1 norm penalty 
fit.sdfm$em$num_iter
#>  [1] 17  6 14  2  2  2  3  3  3  3 18  3 12  5  5

# Optimal L1 norm penalty chosen
fit.sdfm$em$alpha_opt
#> [1] 4.54091
\end{verbatim}

We now explore the estimated factors and loadings for the regular and sparse DFM fits. We are able to group the indicator series into colours depending on the source of the indicator and use the \verb+type = "loading.grouplineplot"+ setting in \verb+plot()+. We set the trade in goods (TiG) target black, IoP blue, CPI red, PPI pink, exchange rate (Exch) green, BCI \& CCI (Conf) navy and google trends (GT) brown. This will make it easier to visualise which indicators are loading onto specific factors. Factor and loading plots can be generated by:
\begin{verbatim}
## Plot the estimated factors for the DFM
plot(fit.dfm, type = 'factor')  

## Plot the estimated factors for the Sparse DFM
plot(fit.sdfm, type = 'factor')

## Specify the name of the group each indicator belongs too
# this is a vector of length matching the number of columns of the data
groups = c(rep('TiG',9), rep('IoP',89), rep('CPI',166), rep('PPI',153),
          rep('Exch',12), rep('Conf',2), rep('GT',14))

## Specify the colours for each of the groups 
# this is a vector of length matching the number of different groups 
group_cols = c('black','blue','red','pink','green','navy','brown')

## Plot the DFM group lineplot in a 2 x 2 grid - uses the gridExtra package from CRAN.
p1 = plot(fit.dfm, type = 'loading.grouplineplot', loading.factor = 1, 
          group.names = groups, group.cols = group_cols)
p2 = plot(fit.dfm, type = 'loading.grouplineplot', loading.factor = 2,
          group.names = groups, group.cols = group_cols)
p3 = plot(fit.dfm, type = 'loading.grouplineplot', loading.factor = 3,
          group.names = groups, group.cols = group_cols)
p4 = plot(fit.dfm, type = 'loading.grouplineplot', loading.factor = 4,
          group.names = groups, group.cols = group_cols)

grid.arrange(p1, p2, p3, p4, nrow = 2)

## Do the same for the Sparse DFM fit 
p5 = plot(fit.sdfm, type = 'loading.grouplineplot', loading.factor = 1,
          group.names = groups, group.cols = group_cols)
p6 = plot(fit.sdfm, type = 'loading.grouplineplot', loading.factor = 2,
          group.names = groups, group.cols = group_cols)
p7 = plot(fit.sdfm, type = 'loading.grouplineplot', loading.factor = 3,
          group.names = groups, group.cols = group_cols)
p8 = plot(fit.sdfm, type = 'loading.grouplineplot', loading.factor = 4,
          group.names = groups, group.cols = group_cols)

grid.arrange(p5, p6, p7, p8, nrow = 2)
\end{verbatim}

\begin{figure}[h!]
    \centering
    \includegraphics[width=0.49\linewidth]{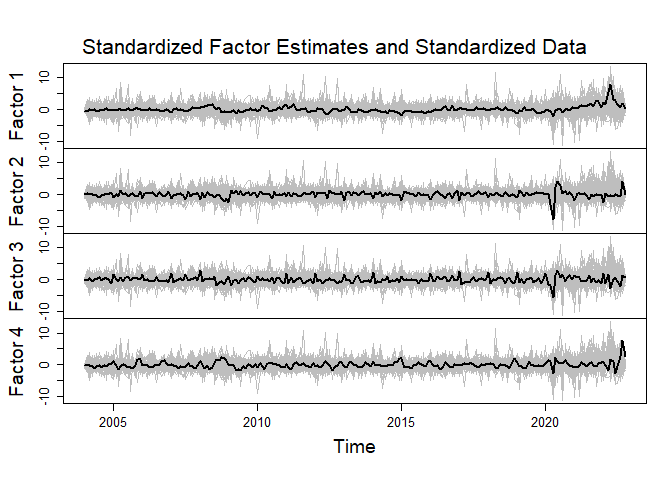}
    \includegraphics[width=0.49\linewidth]{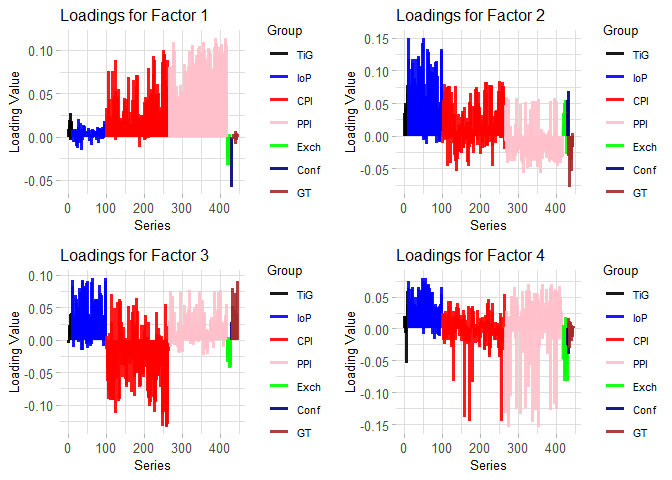} \\
    \includegraphics[width=0.49\linewidth]{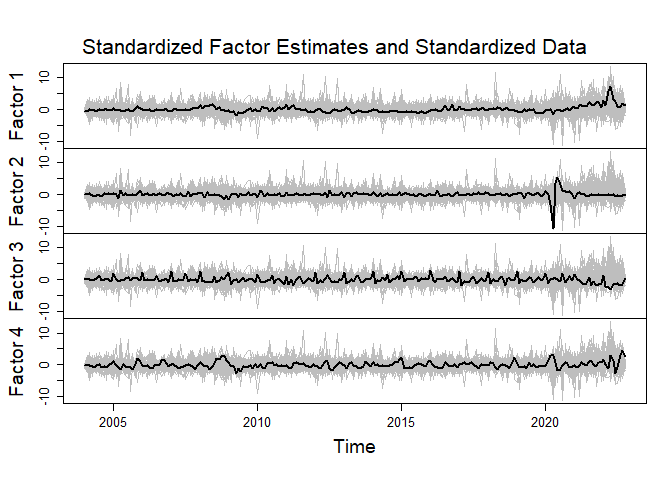}
    \includegraphics[width=0.49\linewidth]{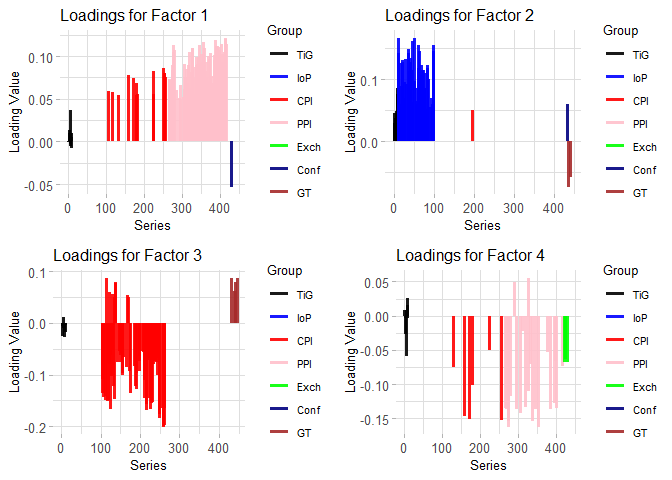} 
    \caption{Estimated factor estimates and corresponding loadings for the exports application with a regular DFM fit on the top row and a Sparse DFM fit on the bottom row. }
    \label{fig:exportsoutputs}
\end{figure}

Figure \ref{fig:exportsoutputs} displays the generated factor and loadings plots for the regular DFM fit on the top row and the Sparse DFM fit on the bottom row. We have used the R package \textbf{gridExtra} to neatly display the loading plots in a $2\times 2$ grid. As all variables are loaded onto all the factors in a regular DFM it is very difficult to interpret the factor structure from these loading plots. The loadings from all groups in every factor are quite large and it is impossible to make conclusions on which data groups are related to specific factors. On the other hand, with sparse factor loadings, we are able to make a lot clearer conclusions on factor structure. We can clearly visualise which indicator series are the driving force between each factor. Factor 2 for example, with the obvious drop in early 2020 due to the COVID pandemic, is heavily loaded with indicators coming from the Index of Production, confidence indices and google trends. Index of Production does not actually appear in any other factors and we can view this as a clear indicator of the COVID drop. It is interesting that google search words to do with trade in goods are present in factor 2 as well. With lots of economic volatility in recent years, using google trends search words may be a very useful indicator of economic activity. Factor 1 seems to be mainly loaded with PPI data, while factor 3 is heavily loaded with CPI data. Some inflation data and exchange rate indicators are present in factor 4, which seems to have shocks during the 2009 and 2020 recessions. Note that in all 4 factor loading plots the trade in goods target series are loaded as we specified q = 9 in the sparseDFM fit to ensure these variables are not regularised.

We now turn our attention to nowcasts of the missing months of September and October 2022. It is very easy to extract nowcasts from the \verb+sparseDFM+ fit. As the data was inputted with the ragged edge structure, with NAs coded in for September and October 2022 for the target series, the sparseDFM output will provide us with estimates for these missing months. There are two ways these values can be extracted:
\begin{verbatim}
## DFM nowcasts (on the differenced data)

# (1) directly from fit.dfm 
dfm.nowcasts = tail(fit.dfm$data$fitted.unscaled[,1:9],2)

# which is the same as

# (2) from fitted()
dfm.nowcasts = tail(fitted(fit.dfm)[,1:9],2)

## Sparse DFM nowcasts (on the differenced data)

sdfm.nowcasts = tail(fit.sdfm$data$fitted.unscaled[,1:9],2)
\end{verbatim}
To transform these first-differenced-nowcasts into nowcasts on the original level data, we need to undifference. To do this we can take the most recent observed value (August 2022) and add the September first-difference-nowcast for the September level nowcast, and then add the October first-difference-nowcast to this value to get the October level nowcast:
\begin{verbatim}
## August 2022 figures for targets 

obs_aug22 = tail(data,3)[1,1:9]

## DFM nowcast for original level 

dfm_sept_nowcast = obs_aug22 + dfm.nowcasts[1,]
dfm_oct_nowcast = dfm_sept_nowcast + dfm.nowcasts[2,]

## Sparse DFM nowcast for original level 

sdfm_sept_nowcast = obs_aug22 + sdfm.nowcasts[1,]
sdfm_oct_nowcast = sdfm_sept_nowcast + sdfm.nowcasts[2,]

# Print 
cbind(dfm_sept_nowcast,
dfm_oct_nowcast,
sdfm_sept_nowcast,
sdfm_oct_nowcast)
#>                                           dfm_sept_nowcast dfm_oct_nowcast
#> target.Food & live animals                      1466.20494      1487.89722
#> target.Beverages and tobacco                     892.52105       910.32919
#> target.Crude materials                           931.82943       935.86939
#> target.Fuels                                    5434.76988      5281.31709
#> target.Animal and vegetable oils and fats         81.11094        82.15963
#> target.Chemicals                                5450.35939      5496.92373
#> target.Material manufactures                    4250.23561      4348.36403
#> target.Machinery and transport                 14966.96739     15394.04752
#> target.Miscellaneous manufactures               4575.18045      4759.21107
#>                                           sdfm_sept_nowcast sdfm_oct_nowcast
#> target.Food & live animals                       1373.09109       1379.51984
#> target.Beverages and tobacco                      806.32814        810.16403
#> target.Crude materials                            843.30349        828.98577
#> target.Fuels                                     5444.14078       5284.05460
#> target.Animal and vegetable oils and fats          79.86784         80.48402
#> target.Chemicals                                 5233.46691       5237.28705
#> target.Material manufactures                     3881.44541       3923.95003
#> target.Machinery and transport                  12776.89494      12788.07209
#> target.Miscellaneous manufactures                3896.18607       3951.78687
\end{verbatim}

To determine which model is producing the most accurate nowcasts we can set up a \textbf{pseudo real-time nowcasting exercise} where we try and predict the missing ragged edge in an \textbf{expanding window from August 2018 to July 2022}. As it would be difficult to store all previous vintages of data back to 2018, we just use the complete data set (up to October 2022) and at each nowcasting window create a ragged edge following the 2 month lag for the targets and IoP and 1 month lag for everything else except google trends. We begin with the window Jan 2004 - Aug 2018, where the current available target exports data is June 2018. We produce nowcasts for July 2018 (representing a nowcast 1 month prior to release) and for August 2018 (representing a nowcast 2 months prior to release). We produce these nowcasts for horizon 1 and 2 for every month up to July 2022 - a total of 48 months. The code for the pseudo real-time nowcasting exercise is presented in Appendix \ref{appendixG}. Note, we fit a new model at each nowcasting window and re-tune for the best $\ell_1$-norm penalty parameter. An alternative could be to store the model parameters estimated from one fit of the SDFM to a particular time window and then just run the \verb+kalmanUnivariate()+ function on these stored parameters at each time window. 

Table \ref{tab:pseudo} displays the average, 25th, 50th and 75th percentiles of the mean absolute error across the 48 month expanding window for horizon 1 and 2 in both models. It appears that Sparse DFM is performing better than a regular DFM. It has a lower average mean absolute error and tighter bands around the median. As expected, the error for horizon 1 is slightly lower than horizon 2 as it is able to exploit all indicators with a 1 month lag in its estimation.

\begin{table}[h!]
    \centering
    \begin{tabular}{c|c|c|c|c|c}
     & & \textbf{Mean} & $\bm{25\%}$ & $\bm{50\%}$ & $\bm{75\%}$ \\ \hline 
     \multirow{2}{*}{\textbf{Horizon 1}} & \textbf{DFM} & 373.72 & 152.403 & 232.856 & 455.077 \\
                                & \textbf{SDFM} & 297.233 & 141.118 & 212.392 & 306.925 \\ \hline 
    \multirow{2}{*}{\textbf{Horizon 2}} & \textbf{DFM} & 437.15 & 201.821 & 302.672 & 515.081 \\
                                & \textbf{SDFM} & 357.602 & 161.937 & 233.790 & 366.392 \\
\end{tabular}
    \caption{Pseudo real-time nowcasting exercise results.}
    \label{tab:pseudo}
\end{table}

\section[Conclusions]{Conclusions}\label{sec5}
The R package \textbf{sparseDFM} has been developed to allow the fitting of various estimation methods for dynamic factor models (DFMs) including PCA \citep{stock2002forecasting}, 2Stage \citep{giannone2008nowcasting}, EM \citep{banbura2014maximum}, and the novel EM-sparse approach for sparse DFMs \citep{mosleySAC}. We implement both multivariate \citep{shumway1982approach}, and univariate \citep{koopman2000fast} Kalman filter and smoother equations in C++ using the Armadillo library \citep{rcpparmadillo}, as well as the ADMM updates for the EM-sparse algorithm for computational speed. 

The implementation of an efficient and interpretable EM framework for sparse DFM estimation is new to the R programming language and is available from the Comprehensive R Archive Network (CRAN) at \url{https://cran.r-project.org/}. The real-data case studies on UK inflation and nowcasting Trade in Goods displayed the benefits of a sparse DFM; from understanding factor structure to nowcasting performance. The \textbf{sparseDFM} package will be maintained at \url{https://github.com/mosleyl/sparseDFM} and we hope add new features in the future including new plotting capabilities and extended assumptions on the underlying VAR(1) latent factor process.

\subsection*{Acknowledgments}
AG and TTC acknowledge funding from the EPSRC grant EP/T025964/1. AG and LM were supported by the ESRC grant ES/V006339/1. LM acknowledges support from the STOR-i Centre for Doctoral Training and the Office for National Statistics. 

\bibliographystyle{chicago}
\bibliography{References}

\newpage

\appendix

\section[Multivariate and Univariate Kalman Filter and Smoother Equations]{Multivariate and Univariate Kalman Filter and Smoother Equations}\label{appendixA}
In this Appendix we provide the classic multivariate Kalman filter and smoother (KFS) equations of \cite{shumway1982approach} and the fast univariate treatment KFS equations of \cite{koopman2000fast} that can be used to estimate the mean and covariance of the latent states in a state space framework (\ref{eq:observation})-(\ref{eq:factor}) based on observations of the system. In a DFM, the latent states are the factors\footnote{Note that in the extension to AR(1) idiosyncratic errors, as discussed in Section \ref{sec2.5}, the latent state involves the factors augmented with the AR(1) errors. The KFS equations in this Appendix are analogous for both cases with the notation $\bF_t$ representing the latent state - either just the factors or the factors augmented with AR(1) errors. Similarly, the parameters can be swapped, $\btheta$ with $\tilde{\btheta}$. I.e. these equations are for a general state-space model of the form (\ref{eq:observation})-(\ref{eq:factor}).}, $\bF_t$. We use the following notation for the conditional mean and covariances of the state:
\begin{align*}
    \ba_{t|s} &= \E[\bF_t|\bomega_s] \, ,\\
    \bP_{t|s} &= \mathrm{Var}[\bF_t|\bomega_s] \, ,\\
    \bP_{t,t-1|s} &= \mathrm{Cov}[\bF_t, \bF_{t-1}|\bomega_s] \, ,
\end{align*}
conditional on all information we have observed up to a time $s$, denoted by $\bomega_s$.

Filtering aims to make recursive predictions of the mean and covariance of the state at $t$ given information up to $t-1$, i.e. finds $\ba_{t|t-1}$ and $\bP_{t|t-1}$ starting from $t=1$ to $t=n$. Smoothing then works backwards in time to find smoothed estimates of the mean and covariance of the state given all information, i.e. finds $\ba_{t|n}$ and $\bP_{t|n}$ starting from $t=n$ to $t=1$. 

\subsection[Multivariate KFS Equations - Shumway and Stoffer (1982)]{Multivariate KFS Equations - \cite{shumway1982approach}}
The classic approach to filtering and smoothing found in \cite{shumway1982approach} and detailed in \cite{harvey1990forecasting} is based on considering the contribution of every variable in the observational vector at each successive time point. We therefore describe this as a multivariate approach and set the input in \verb+SparseDFM()+ as \verb+kalman = multivariate+.

Assuming the set of parameters $\btheta = (\blam, \bA, \bsig_{\beps}, \bsig_{\bu})$ and initial conditions $\ba_{0|0}$ and $\bP_{0|0}$ are known, the filtering equations for $t=1,\dots,n$ are
\begin{align*}
    \ba_{t|t-1} &= \bA\ba_{t-1|t-1} \tag*{(predicted factor estimate)} \\
    \bP_{t|t-1} &= \bA\bP_{t-1|t-1}\bA^\top + \bsig_u \tag*{(predicted factor covariance)} \\
    \bv_t &= \bX_t - \blam \ba_{t|t-1} \tag*{(innovation error)} \\
    \bC_t &= \blam \bP_{t|t-1} \blam^\top + \bsig_{\beps} \tag*{(innovation covariance)} \\
    \bK_t &= \bP_{t|t-1} \blam^\top \bC_t^{-1} \tag*{(Kalman gain)} \\
    \ba_{t|t} &= \ba_{t|t-1} + \bK_t\bv_t \tag*{(updated factor estimate)} \\
    \bP_{t|t} &= \bP_{t|t-1} - \bK_t \blam \bP_{t|t-1} \tag*{(updated factor covariance)}
\end{align*}
and the smoothing equations for $t=n,\dots,1$ are
\begin{align*}
    \bJ_{t-1} &= \bP_{t-1|t-1}\bA^\top (\bP_{t|t-1})^{-1} \tag*{(smoother gain)} \\
    \ba_{t-1|n} &= \ba_{t-1|t-1} + \bJ_{t-1}(\bF_{t|n} - \bA \ba_{t-1|t-1}) \tag*{(smoothed factor estimate)} \\
    \bP_{t-1|n} &= \bP_{t-1|t-1} + \bJ_{t-1}(\bP_{t|n} - \bP_{t|t-1})\bJ_{t-1}^\top \tag*{(smoothed factor covariance)}
\end{align*}
where we initialise $\ba_{n|n}$ and $\bP_{n|n}$ based on the last filter estimates. At any stage where an observation $X_{t,i}$ is missing at time $t$, we omit this variable from the calculation of the Kalman gain $\bK_t$. The  lagged-covariance matrix $\bP_{t,t-1|n}$ is found using backwards recursions on:
\begin{equation}
    \bP_{t-1,t-2|n} = \bP_{t-1|t-1}\bJ_{t-2}^\top + \bJ_{t-1}(\bP_{t,t-1|n} - \bA\bP_{t-1|t-1})\bJ_{t-2}^\top \, ,
    \label{vvsmooth}
\end{equation}
for $t=n,\dots,2$, where 
\begin{equation*}
    \bP_{n,n-1|n} = (\bI - \bK_n\blam)\bA\bP_{n-1|n-1} \, .
\end{equation*}

The filtering equations require the covariance of innovation errors $\bC_t$ to be inverted. This can be difficult when $p$ is large. \cite{harvey1990forecasting} show this inversion can be made easier by making use of the Woodbury identity and find:
\begin{equation}
    \bC_t^{-1} = \bsig_e^{-1} - \bsig_e^{-1}\blam(\bP_{t|t-1}^{-1} + \blam^\top \bsig_e^{-1}\blam)^{-1}\blam^\top \bsig_{\beps}^{-1} \, ,
    \label{cinv}
\end{equation}
which is valid when $\bsig_{\beps}$ and $\bP_{t|t-1}$ are non-singular.
This inversion is easy to evaluate if $\bsig_e$ is diagonal (as assumed in exact factors models). In cases when $p$ is much larger than $r$, as is the case in the majority of DFM problems, the inversion of $\bP_{t|t-1}$ is a much easier operation than the inversion of $\bC_t$ directly. 

\cite{harvey1990forecasting} further show that the determinant of $\bC_t$ can be written as:
\begin{equation}
    |\bC_t| = |\bsig_{\beps}| \times |\bP_{t|t-1}| \times |\bP_{t|t-1}^{-1} + \blam^\top \bsig_{\beps}^{-1}\blam| \, .
    \label{cdet}
\end{equation}
Both (\ref{cinv}) and (\ref{cdet}) are used in the calculation of the log-likelihood of the innovations:
\begin{equation}
    \log\mathcal{L}(\bX ; \btheta) =  - \frac{1}{2}\sum_{t=1}^n \left(p_t\log(2\pi) + \log |\bC_t| + \bv_t \bC_t^{-1}\bv_t \right)\, .
    \label{llinv}
\end{equation}
where $p_t$ is the number of variables observed (not-missing) at $t$ and $\btheta$ is the collection of all parameters in (\ref{eq:measurement})-(\ref{eq:state}). Note,
this log-likelihood is used to determine convergence of the EM algorithm as in (\ref{eq:loglconverge}) of the EM approaches in Sections \ref{sec2.3} and \ref{sec2.4}.

\subsection[Univariate KFS Equations - Koopman and Durbin (2000)]{Univariate KFS Equations - \cite{koopman2000fast}}

The classic KFS equations can be slow when $p$ is large. Since we assume $\bsig_{\beps}$ is diagonal, we can equivalently filter the observations $\bX_t$ one element at a time, as opposed to all together as in the classic approach \citep{durbin2012time, koopman2000fast}. As matrix inversion becomes scalar divisions, huge speedups are possible. The input to the \verb+SparseDFM()+ function is \verb+kalman = univariate+.

Let us define the individual elements $\bX_t = (X_{t,1},\dots,X_{t,p})^\top$, $\blam = (\blam_1^\top,\dots,\blam_p^\top)^\top$, $\bsig_{\beps} = diag(\sigma_{\epsilon 1}^2,\dots,\sigma_{\epsilon p}^2)$ and define new notation similar to that found in \cite{durbin2012time}:
\begin{align*}
\ba_{t,i} &= \E[\bF_{t,i}|\bomega_{t-1},X_{t,1}, \dots, X_{t,i-1}] \, , \\
\ba_{t,1} &= \E[\bF_{t,1}|\bomega_{t-1}] \, , \\
\bP_{t,i} &= \mathrm{Var}[\bF_{t,i}|\bomega_{t-1},X_{t,1}, \dots, X_{t,i-1}] \, , \\
\bP_{t,1} &= \mathrm{Var}[\bF_{t,1}|\bomega_{t-1}] \, ,
\end{align*}
for $i=1,\dots,p$ and $t=1,\dots,n$.

The univariate treatment \citep{koopman2000fast} iterates the below equations. These update equations are equivalent in form to the multivariate ones of the classic \cite{shumway1982approach} approach except that the $t$ subscript now becomes a $t,i$ subscript, and the $t|t$ subscript now becomes $t,i+1$. This means that we are filtering spatiotemporally, processing one $i$ after another in a sequential manner, instead of updating all $p$ measurements at once:
\begin{align*}
v_{t,i} &= X_{t,i}-\blam_{i}\ba_{t,i}, \\
C_{t,i} &= \blam_{i}\bP_{t,i}\blam_{i}^\top+\sigma^2_{\epsilon,i}, \\
\bK_{t,i} &= \bP_{t,i}\blam_{i}^\top C_{t,i}^{-1}, \\
\ba_{t,i+1} &= \ba_{t,i}+\bK_{t,i}v_{t,i}, \\
\bP_{t,i+1} &= \bP_{t,i}-\bK_{t,i}C_{t,i}\bK_{t,i}^\top,
\end{align*}
for $i=1,\ldots,p$ and $t=1,\ldots,n$. If $X_{t,i}$ is missing or $C_{t,i}$ is zero, omit the term containing $\bK_{t,i}$. The transition to $t+1$ is given by the following prediction equations:
\begin{align*}
\ba_{t+1,1} &= \bA \ba_{t,p+1}, \\
\bP_{t+1,1} &= \bA \bP_{t,p+1}\bA^\top+\bsig_{\bu}.
\end{align*}
These prediction equations are exactly the same as the multivariate ones (i.e., predictions are not treated sequentially but all at once). From our perspective, univariate treatment is better called univariate updates plus multivariate predictions.

Unlike \cite{shumway1982approach}, the measurement update comes before the transition; however, we can revert to doing the transition first if our initial state means and covariances start from $t=0$ instead of $t=1$. Likewise, univariate smoothing is defined by:
\begin{align*}
\bL_{t,i} &= \bI_m - \bK_{t,i} \blam_{i}, \\
\br_{t,i-1} &= \blam_{i}^\top C_{t,i}^{-1}v_{t,i}+\bL_{t,i}^\top \br_{t,i}, \\
\bN_{t,i-1} &= \blam_{i}^\top C_{t,i}^{-1}\blam_{i}+\bL_{t,i}^\top\bN_{t,i}\bL_{t,i}, \\
\br_{t-1,p} &= \bA^\top \br_{t,0}, \\
\bN_{t-1,p} &= \bA^\top \bN_{t,0}\bA,
\end{align*}
for $i=p,\ldots,1$ and $t=n,\ldots,1$, with $\br_{n,p}$ and $\bN_{n,p}$ initialised to $0$. Again, if $X_{t,i}$ is missing or $C_{t,i}$ is zero, drop the terms containing $\bK_{t,i}$. Finally, the equations for $\ba_{t|n}$ and $\bP_{t|n}$ are:
\begin{align*}
\ba_{t|n} &= \ba_{t,1}+\bP_{t,1}\br_{t,0}, \\
\bP_{t|n} &= \bP_{t,1}-\bP_{t,1}\bN_{t,0}\bP_{t,1}.
\end{align*}
These results will be equivalent to $\ba_{t|n}$ and $\bP_{t|n}$ from the classic multivariate approach, yet obtained with substantial improvement in computational efficiency. 

In order to calculate the cross-covariance matrix $\bP_{t,t-1|n}$, we use \cite{jong1988covariances}'s theorem:
\begin{equation*}
\bP_{t-1,t|n} = \bP_{t-1|t-1}\bA^\top (\bP_{t|t-1})^{-1}\bP_{t|n},
\end{equation*}
which, after transposition, gives
\begin{equation*}
\bP_{t,t-1|n} = \bP_{t|n}(\bP_{t|t-1})^{-1}\bA \bP_{t-1|t-1}.
\end{equation*}

The log-likelihood for the above filter \citep{durbin2012time, koopman2000fast} is given below:
\begin{align}
\log\mathcal{L}(\bX ; \btheta) &= -\frac{1}{2}\sum_{t=1}^{n}\sum_{i=1}^{p}\iota_{t,i}\left(\log(2\pi)+\log |C_{t,i}|+v_{t,i}^\top C_{t,i}^{-1}v_{t,i}\right) \\
&= -\frac{1}{2}\sum_{t=1}^{n}\sum_{i=1}^{p}\iota_{t,i}\left(\log(2\pi)+\log C_{t,i}+\frac{v_{t,i}^{2}}{C_{t,i}}\right),
\end{align}
as $C_{t,i}$ and $v_{t,i}$ are scalars. Here, $\iota_{t,i}$ is zero if $X_{t,i}$ is missing or $C_{t,i}$ is zero, or one otherwise.

\section{EM Algorithm Derivations for Section 2.3}\label{appendixB}
The EM algorithm of \cite{banbura2014maximum} involves calculating and maximising the expected log-likelihood of the DFM conditional on the available information $\bomega_n$. Given the log-likelihood in (\ref{eq:jointlogl}), the conditional expected log-likelihood is 
\begin{align}
     \E \left[\log\mathcal{L}(\bX,\bF;\btheta)\vert\bomega_n\right]  =& -\frac{1}{2}\log \vert\mathcal{\bP}_0\vert
     - \mathrm{tr}\left\{\mathcal{\bP}_0^{-1} \E \left[(\bF_0 - \bm{\alpha}_0)(\bF_0 - \bm{\alpha}_0)^\top \vert\bomega_n\right] \right\}  \label{eq:Einit}\\
     & - \frac{n}{2} \log \vert\bsig_{\bu}\vert - \frac{1}{2}\sum_{t=1}^n \mathrm{tr}\left\{ \bsig_{\bu}^{-1} \E \left[\bu_t\bu_t^\top \vert \bomega_n\right]\right\} \label{eq:Estate} \\
     & - \frac{n}{2} \log \vert\bsig_{\beps}\vert - \frac{1}{2}\sum_{t=1}^n \mathrm{tr}\left\{ \bsig_{\beps}^{-1} \E\left[\beps_t\beps_t^\top \vert \bomega_n\right] \right\} \label{eq:Emeas}\; . 
\end{align}
We refer to $\E \left[\log\mathcal{L}(\bX,\bF;\btheta)|\bomega_n \right ] $ just as $\mathcal{E}$ to make notation easier. In each derivation we use the linearity of conditional expectation: $\mathrm{Cov}(X,Y|A)=\E(XY|A)-\E(X|A)\E(Y|A)$ for two random variables $X,Y$ and non-zero event $A$. We also use the derivative property $\partial  (\log |\bZ|)/\partial  (\bZ^{-1}) = -\bZ$ for a matrix $\bZ$. Recall, we define the conditional mean and covariances of the state as:
\begin{align*}
    \ba_{t|n} &= \E[\bF_t|\bomega_n] \, ,\\
    \bP_{t|n} &= \mathrm{Var}[\bF_t|\bomega_n] \, ,\\
    \bP_{t,t-1|n} &= \mathrm{Cov}[\bF_t, \bF_{t-1}|\bomega_n] \, ,
\end{align*}
conditional on all information we have observed up to $n$, denoted by $\bomega_n$, and let $\bS_{t\vert n} = \ba_{t\vert n}\ba_{t\vert n}^{\top} + \bP_{t\vert n}$, and $\bS_{t,t-1\vert n} = \ba_{t\vert n}\ba_{t-1\vert n}^{\top} + \bP_{t,t-1\vert n}$.

Expanding out the conditional expectation in the initial state distribution part (\ref{eq:Einit}) we obtain:
\begin{align*}
    \E[(\bF_0 - \bm{\alpha}_0)(\bF_0 - \bm{\alpha}_0)^\top |\bomega_n ] &= \E[\bF_0\bF_0^\top - 2\bF_0\bm{\alpha}_0^\top + \bm{\alpha}_0\bm{\alpha}_0^\top |\bomega_n ]  \\
    &= \E[\bF_0\bF_0^\top | \bomega_n] - 2\E[\bF_0|\bomega_n]\bm{\alpha}_0^\top + \bm{\alpha}_0\bm{\alpha}_0^\top  \\
    &= \E[\bF_0 | \bomega_n]\E[\bF_0^\top | \bomega_n] + \Var[\bF_0 | \bomega_n] - 2\E[\bF_0|\bomega_n]\bm{\alpha}_0^\top + \bm{\alpha}_0\bm{\alpha}_0^\top \\
    &= \ba_{0\vert n}\ba_{0\vert n}^\top + \bP_{0 \vert n} - 2\ba_{0\vert n}\bm{\alpha}_0^\top + \bm{\alpha}_0\bm{\alpha}_0^\top = \bS_{0\vert n}-2\ba_{0\vert n}\bm{\alpha}_0^\top + \bm{\alpha}_0\bm{\alpha}_0^\top \, .
\end{align*}
Differentiating $\mathcal{E}$ with respect to $\bm{\alpha}_0$ and $\mathcal{\bP}_0^{-1}$ we obtain:
\begin{align*}
    \frac{\partial \mathcal{E}}{\partial \bm{\alpha}_0} &=  - \frac{1}{2}\mathcal{\bP}_0^{-1}\frac{\partial}{\partial \bm{\alpha}_0}\E[(\bF_0 - \bm{\alpha}_0)(\bF_0 - \bm{\alpha}_0)^\top |\bomega_n ] = - \frac{1}{2}\mathcal{\bP}_0^{-1}\left( -2\ba_{0\vert n} + 2\bm{\alpha}_0 \right) \\
    \frac{\partial \mathcal{E}}{\partial \mathcal{\bP}_0^{-1}} &= \frac{1}{2}\mathcal{\bP}_0 - \frac{1}{2}\E[(\bF_0 - \bm{\alpha}_0)(\bF_0 - \bm{\alpha}_0)^\top |\bomega_n ] \\
    &= \frac{1}{2}\mathcal{\bP}_0 - \frac{1}{2}\left( \bS_{0\vert n} - 2\ba_{0\vert n}\bm{\alpha}_0^\top + \bm{\alpha}_0\bm{\alpha}_0^\top\right) \, .
\end{align*}
Setting both derivatives equal to 0, we first get the result
\begin{equation*}
    \hat{\bm{\alpha}}_0 = \ba_{0\vert n} \, ,  
\end{equation*}
and use this to obtain 
\begin{equation*}
    \hat{\mathcal{\bP}}_0 =\bP_{0\vert n}  \,.
\end{equation*}

Expanding out the expectation in the state transition part (\ref{eq:Estate}) we obtain:
\begin{align*}
    \E \left[\bu_t^\top \bu_t \vert \bomega_n\right] &= \E [(\bF_t - \bA \bF_{t-1})(\bF_t - \bA \bF_{t-1})^\top |\bomega_n ] 
    = 
    \E[ \bF_t\bF_t^\top - 2\bF_t\bF_{t-1}^\top\bA^\top + \bA\bF_{t-1}\bF_{t-1}^\top\bA^\top | \bomega_n]  \\
    &= \E[\bF_t\bF_t^\top|\bomega_n] - 2\E[\bF_t\bF_{t-1}^\top|\bomega_n]\bA^\top + \bA\E[\bF_{t-1}\bF_{t-1}^\top|\bomega_n]\bA^\top  \\
    &= \E[\bF_t | \bomega_n]\E[\bF_t^\top | \bomega_n] + \Var[\bF_t | \bomega_n]  
      - 2\left( \E[\bF_t|\bomega_n]\E[\bF_{t-1}^\top|\bomega_n] + \mathrm{Cov}[\bF_t,\bF_{t-1}^\top|\bomega_n]\right)\bA^\top  \\
    & \quad + \bA\left( \E[\bF_{t-1}|\bomega_n]\E[\bF_{t-1}^\top|\bomega_n]+\mathrm{Var}[\bF_{t-1}|\bomega_n]\right)\bA^\top \\
    &= \bS_{t\vert n} - 2\bS_{t,t-1 \vert n}\bA^\top + \bA\bS_{t-1 \vert n}\bA^\top \, .
\end{align*}

Differentiating $\mathcal{E}$ with respect to $\bA$ and $\bsig_{\bu}^{-1}$ we obtain:
\begin{align*}
    \frac{\partial \mathcal{E}}{\partial \bA} &=  - \frac{1}{2}\bsig_{\bu}^{-1}\frac{\partial}{\partial \bA} \sum_{t=1}^n\E \left[\bu_t^\top \bu_t \vert \bomega_n\right] 
    = -\frac{1}{2}\bsig_{\bu}^{-1}\sum_{t=1}^n\left( -2\bS_{t,t-1\vert n} + 2\bA\bS_{t-1\vert n}\right) \, , \\
    \frac{\partial \mathcal{E}}{\partial \bsig_{\bu}^{-1}} &=  \frac{n}{2}\bsig_{\bu}- \frac{1}{2}\sum_{t=1}^n\E \left[\bu_t^\top \bu_t \vert \bomega_n\right] = \frac{n}{2}\bsig_{\bu}- \frac{1}{2}\left( \bS_{t\vert n} - 2\bS_{t,t-1 \vert n}\bA^\top + \bA\bS_{t-1 \vert n}\bA^\top \right) \, .
\end{align*}
Setting both derivatives equal to 0, we first get the result
\begin{equation*}
    \hat{\bA} = \left(\sum_{t=1}^n \bS_{t-1\vert n}\right)^{-1} \left(\sum_{t=1}^n \bS_{t,t-1\vert n}\right) \, ,
\end{equation*}
and use this to obtain 
\begin{equation*}
    \hat{\bsig}_{\bu} = \frac{1}{n} \sum_{t=1}^n\left[\bS_{t\vert n} - \hat{\bA} \left( \bS_{t-1,t\vert n}  \right) \right] \, .
\end{equation*}

Using the selection matrix $\bW_t$ defined in Section \ref{sec2.3}, with the property $\bX_t = \bW_t\bX_t + (\bI-\bW_t)\bX_t $, we can expand out the expectation in the measurement equation part (\ref{eq:Emeas}) and obtain:
\begin{align*}
    & \E\left[\beps_t\beps_t^\top \vert \bomega_n\right] = \E[(\bX_t - \blam \bF_t)(\bX_t - \blam \bF_t)^\top  |\bomega_n ]  \\
    &= \E [ [ \bW_t(\bX_t-\blam\bF_t)+(\bI-\bW_t)(\bX_t-\blam\bF_t)] [ \bW_t(\bX_t-\blam\bF_t)+(\bI-\bW_t)(\bX_t-\blam\bF_t)]^\top |\bomega_n ] \\
    &= \E [\bW_t(\bX_t-\blam\bF_t)(\bX_t-\blam\bF_t)\bW_t + (\bI-\bW_t)(\bX_t-\blam\bF_t)(\bX_t-\blam\bF_t)^\top\bW_t \\ 
    &\quad + \bW_t(\bX_t - \blam\bF_t)(\bX_t-\blam\bF_t)^\top(\bI-\bW_t) + (\bI-\bW_t)(\bX_t - \blam\bF_t)(\bX_t-\blam\bF_t)^\top(\bI-\bW_t)  |\bomega_n ] \, .
\end{align*}
Using the law of iterated expectations:
\begin{equation*}
    \E [(\bX_t - \blam \bF_t)(\bX_t - \blam \bF_t)^\top  |\bomega_n ]  = \E\left[ \E [(\bX_t - \blam \bF_t)(\bX_t - \blam \bF_t)^\top  |\bF, \bomega_n ] | \bomega_n \right] \, ,
\end{equation*}
\cite{banbura2014maximum} show 
\begin{align*}
    & \E[\bW_t(\bX_t - \blam\bF_t)(\bX_t-\blam\bF_t)^\top(\bI-\bW_t)|\bF,\bomega_n ] = 0\, ,\\
    & \E[(\bI-\bW_t)(\bX_t - \blam\bF_t)(\bX_t-\blam\bF_t)^\top(\bI-\bW_t)|\bF,\bomega_n ] = (\bI-\bW_t)\bsig_{\beps}(\bI-\bW_t) \, , 
\end{align*}
and
\begin{align*}
    \E[\bW_t(\bX_t - \blam\bF_t)(\bX_t-\blam\bF_t)^\top\bW_t|\bomega_n ] 
    &= \bW_t\bX_t\bX_t^\top\bW_t - \bW_t\bX_t\E[\bF_t^\top|\bomega_n]\blam^\top\bW_t \\
    & \quad - \bW_t\blam\E[\bF_t|\bomega_n]\bX_t^\top\bW_t + \bW_t\blam\E[\bF_t\bF_t^\top|\bomega_n]\blam^\top\bW_t \, .
\end{align*}
Hence, we get the result 
\begin{align}
    \E\left[\beps_t\beps_t^\top \vert \bomega_n\right] &= \E[(\bX_t - \blam \bF_t)(\bX_t - \blam \bF_t)^\top  |\bomega_n ] \nonumber \\  
    &= \bW_t\bX_t\bX_t^\top\bW_t - \bW_t\bX_t\E[\bF_t^\top|\bomega_n]\blam^\top\bW_t 
 - \bW_t\blam\E[\bF_t|\bomega_n]\bX_t^\top\bW_t \nonumber \\
    & \quad + \bW_t\blam\left(\E[\bF_t|\bomega_n]\E[\bF_t^\top|\bomega_n]+\mathrm{Var}[\bF_t|\bomega_n] \right)\blam^\top\bW_t + (\bI-\bW_t)\bsig_e^*(\bI-\bW_t) \nonumber \\
    &= \bW_t\bX_t\bX_t^\top\bW_t - 2\bW_t\bX_t\ba_{t\vert n}\blam^\top\bW_t + \bW_t\blam\bS_{t\vert n}\blam^\top\bW_t + (\bI-\bW_t)\bsig_e(\bI-\bW_t) \nonumber \\
    &= \bW_t\left(\bX_t\bX_t^\top - 2\bX_t\ba_{t\vert n}\blam^\top + \blam\bS_{t\vert n}\blam^\top\right)\bW_t + (\bI-\bW_t)\bsig_e^*(\bI-\bW_t) \, ,
    \label{eq:measurementW}
\end{align}
where $\bsig_{\beps}^*$ is obtained from the previous EM
iteration.

Differentiating $\mathcal{E}$ with respect to $\blam$ we obtain:
\begin{equation}
    \frac{\partial \mathcal{E}}{\partial \blam} = -\frac{1}{2}\bsig_{\beps}^{-1}\frac{\partial}{\partial \blam}\sum_{t=1}^n \E\left[\beps_t\beps_t^\top \vert \bomega_n\right] = -\frac{1}{2}\bsig_{\beps}^{-1}\sum_{t=1}^n \bW_t \left( -2\bX_t\ba_{t\vert n} + 2\blam \bS_{t\vert n}\right)\bW_t \, . \label{eq:lamdaderiv}
\end{equation}
Setting this equal to 0, dividing through by $\bW_t\bsig_{\beps}^{-1}$, and using the identity $\mathrm{vec}(ABC)=(C^\top \otimes A)\,\mathrm{vec}(B)$, for matrices $A,B$ and $C$, we obtain the result
\begin{equation*}
  \mathrm{vec}(\hat{\blam}) = \left( \sum_{t=1}^n \bS_{t\vert n} \otimes \bW_t\right)^{-1} \,\mathrm{vec}\left( \sum_{t=1}^n \bW_t\bX_t\ba_{t|n}^\top\right) \, .
\end{equation*}

Differentiating $\mathcal{E}$ with respect to $\bsig_{\beps}^{-1}$ we obtain:
\begin{equation*}
    \frac{\partial \mathcal{E}}{\partial \bsig_{\beps}^{-1}} = \frac{n}{2}\bsig_{\beps} - \frac{1}{2}\sum_{t=1}^n\bW_t\left(\bX_t\bX_t^\top - 2\bX_t\ba_{t\vert n}\blam^\top + \blam\bS_{t\vert n}\blam^\top\right)\bW_t + (\bI-\bW_t)\bsig_e^*(\bI-\bW_t) \, .
\end{equation*}
Setting this equal to 0 and subbing in $\hat{\blam}$ we obtain the result
\begin{equation*}
    \hat{\bsig}_{\beps} = \frac{1}{n}\sum_{t=1}^n \mathrm{diag}\Bigg[  \bW_t \bigg( \bX_t\bX_t^\top\ - 2\bX_t \ba_{t\vert n}^\top \hat{\blam}^\top 
     + \hat{\blam}\bS_{t\vert n}\hat{\blam}^\top\bigg)\bW_t + (\bI - \bW_t)\bsig_{\beps}^*(\bI - \bW_t) \Bigg] \, .
\end{equation*}

\section{ADMM Algorithm Derivations for Section 2.4}\label{appendixC}
For the ADMM algorithm of Section \ref{sec2.3}, used to incorporate $\ell_1$ regularisation into the estimation of the loadings parameter $\blam$, we consider the penalised and augmented Lagrangian
\begin{equation}
    \mathcal{C}(\blam, \bZ, \bU) := -\E\left[\log\mathcal{L}(\bX,\bF;\btheta)|\bomega_n\right] +\alpha\|\bZ\|_{1}+\frac{\nu}{2}\|\blam-\bZ+\bU\|_{F}^{2} \, , \label{eq:lagrangian}
\end{equation}
where $\bZ\in\mathbb{R}^{p\times r}$ is an auxiliary variable, $\bU\in\mathbb{R}^{p\times r}$
are the (scaled) Lagrange multipliers and $\nu$ is the scaling term.
The derivation for minimising with respect to $\bZ$ and $\bU$ are standard and can be found in \cite{boyd2011distributed}. We now go into detail on the minimisation with respect to $\blam$ for iteration $(k+1)$ of the ADMM algorithm:
\begin{equation*}
    \blam^{(k+1)} = \argmin_{\blam} \mathcal{C}(\blam, \bZ^{(k)}, \bU^{(k)}) \, .
\end{equation*}
We can re-write the Lagrangian (\ref{eq:lagrangian}) as
\begin{equation}
    \mathcal{C}(\blam, \bZ, \bU) \propto -\E\left[\log\mathcal{L}(\bX,\bF;\btheta)|\bomega_n\right] +\frac{\nu}{2}\,\mathrm{tr}\left( \blam\blam^\top - 2(\bZ^{(k)} - \bu^{(k)})\blam^\top\right) \, ,
\end{equation}
by dropping terms not involving $\blam$ and using the property $\mathrm{tr}(AB)=\mathrm{tr}(BA)$ for matrices $A$ and $B$. Taking the derivative and using (\ref{eq:lamdaderiv}) we obtain 
\begin{align*}
    \frac{\partial \mathcal{C}(\blam, \bZ^{(k)}, \bU^{(k)})}{\partial \blam} &=  -\frac{\partial}{\partial \blam}\E\left[\log\mathcal{L}(\bX,\bF;\btheta)|\bomega_n\right] + \frac{\nu}{2}\,\mathrm{tr}\left( 2\blam - 2(\bZ^{(k)} - \bu^{(k)})\right) \\
    &= \frac{1}{2}\bsig_{\beps}^{-1}\sum_{t=1}^n \bW_t \left( -2\bX_t\ba_{t\vert n} + 2\blam \bS_{t\vert n}\right)\bW_t + \frac{\nu}{2}\,\mathrm{tr}\left( 2\blam - 2(\bZ^{(k)} - \bu^{(k)})\right) \, .
\end{align*}
 Setting this equal to 0 and re-arranging for $\blam$ we have
\begin{equation*}
    \sum_{t=1}^n \bW_t\bsig_{\beps}^{-1}\bW_t \blam \bS_{t\vert n} + \nu \blam = \sum_{t=1}^n\bW_t\bsig_{\beps}^{-1}\bW_t\bX_t\ba_{t\vert n} + \nu(\bZ^{(k)}-\bU^{(k)}) \, .
\end{equation*}
Using the identity $\mathrm{vec}(ABC)=(C^\top \otimes A)\,\mathrm{vec}(B)$, for matrices $A,B$ and $C$, we obtain the result
\begin{equation*}
\mathrm{vec}(\blam^{(k+1)}) = \left( \sum_{t=1}^n \bS_{t\vert n} \otimes \bW_t\bsig_{\beps}^{-1}\bW_t + \nu\bI_{pr} \right)^{-1}  
\mathrm{vec}\bigg[ \sum_{t=1}^n\bW_t\bsig_{\beps}^{-1}\bW_t\bX_t \ba_{t\vert n}^\top 
+  \nu(\bZ^{(k)}-\bU^{(k)})\bigg] \, .
\end{equation*}

\section{Table to Summarise Generic Functions for S3 Object of Class ``sparseDFM"}\label{appendixD}
\begin{center}
    \begin{tabular}{|m{4cm}|m{11cm}|}
\hline 
    \textbf{Function} & \textbf{Brief Description} \\ \hline 
    \verb+print()+ & Print out the inputs to the \verb+sparseDFM()+ call. \\ \hline
    \verb+summary()+ & Brief summary information on the estimation results for the factor parameters. \\ \hline 
    \verb+plot()+ & Various types of plots available: 
    \begin{itemize}
        \item \verb+factor+ - plot factor estimate series on top of the original standardized stationary data.
        \item \verb+loading.heatmap+ - make a heatmap of the loadings matrix.
        \item \verb+loading.lineplot+ - make a lineplot of variable loadings for a given factor.
        \item \verb+loading.grouplineplot+ - separate variable groups into colours for better visualisation.
        \item \verb+residual+ - boxplot or scatterplot of residuals.
        \item \verb+lasso.bic+ - BIC values for the LASSO tuning parameter.
        \item \verb+em.convergence+ - log-likelihood convergence of EM iterations.
    \end{itemize} \\ \hline
    \verb+fitted()+ & Fitted values of the \verb+sparseDFM()+ fit, i.e. $\hat{\blam}\hat{\bF}$. \\ \hline
    \verb+residuals()/resids()+ & Residual values of the \verb+sparseDFM()+ fit, i.e. $\tilde{\bX} - \hat{\blam}\hat{\bF}$ where $\tilde{\bX}$ is the balanced (nothing-missing) data. \\ \hline
    \verb+predict()+ & Predict the next h steps ahead for the factor estimates and the data series. Returns S3 object of class ``sparseDFM\_forecast". Use \verb+print()+ to print results. \\ \hline
    \end{tabular}
\end{center}

\newpage 

\section{Plots for Section 4.1 - Exploring UK Inflation}\label{appendixE}

\begin{figure}[!htb]
    \centering
    \begin{minipage}{.45\textwidth}
      \centering
        \includegraphics[width=\linewidth]{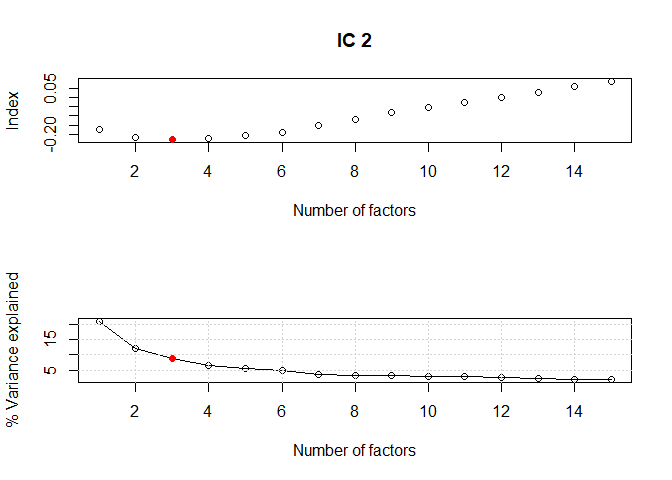}
        \caption{Output of \emph{tuneFactors()} showing the best number of factors to use for the inflation data set.}
         \label{fig:tunefactors}
         \vspace{2em}
    \end{minipage}\hspace{0.5em}
    \begin{minipage}{0.45\textwidth}
        \centering
        \includegraphics[width=\linewidth]{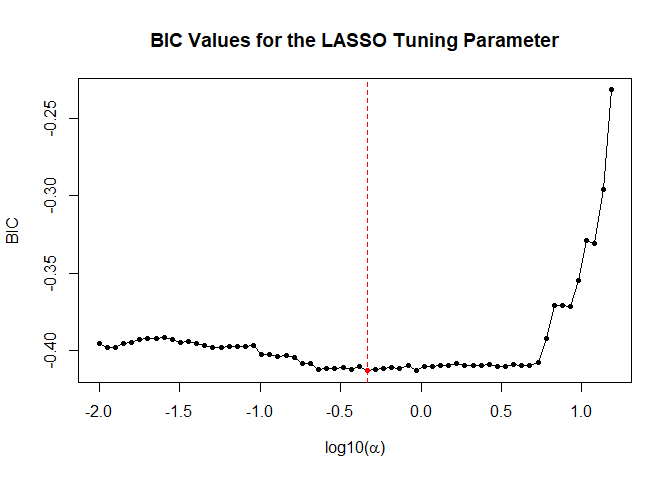}
        \caption{Output of \emph{plot()} with \emph{type = ``lasso.bic"} showing BIC values for $\alpha$'s considered.}
        \label{fig:biccurve}
        \vspace{2em}
    \end{minipage}
    \centering
    \begin{minipage}{.45\textwidth}
      \centering
        \includegraphics[width=\linewidth]{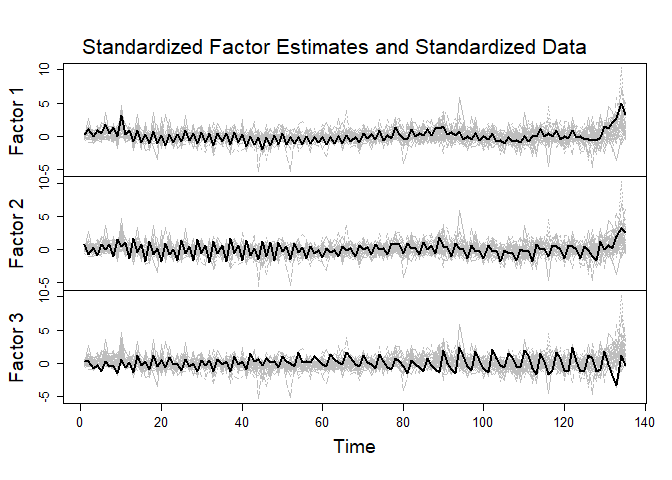}
        \caption{Output of \emph{plot()} with \emph{type = ``factor"} showing the 3 estimated factors on top of the standardised stationary inflation data.}
        \label{fig:factorest}
    \end{minipage}\hspace{0.5em}
    \begin{minipage}{0.45\textwidth}
        \centering
        \includegraphics[width=\linewidth]{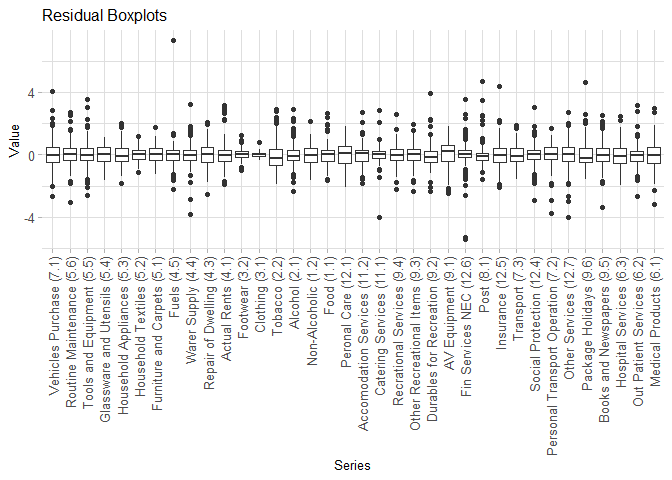}
        \caption{Output of \emph{plot()} with \emph{type = ``residual"} showing the residuals for each inflation series.}
        \label{fig:residualest}
    \end{minipage}
\end{figure}

\newpage 

\section{Plots for Section 4.2 - Nowcasting UK Trade in Goods (Exports)}\label{appendixF}

\begin{figure}[!htb]
    \centering
    \begin{minipage}{.45\textwidth}
       \centering
        \includegraphics[width=\linewidth]{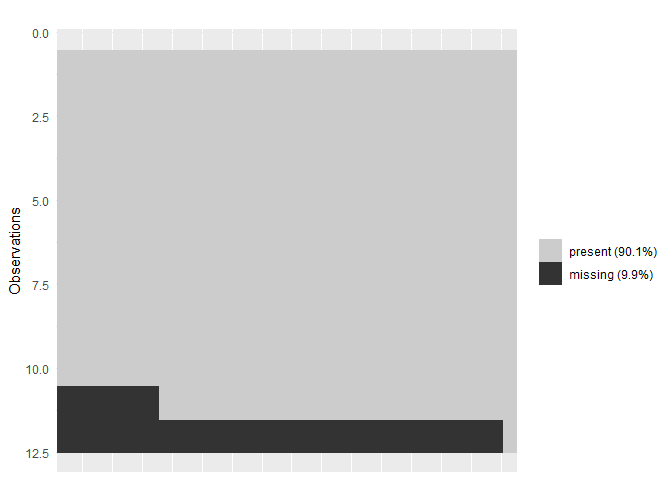}
        \caption{Missing data plot for the last 12 months of the exports full data set.}
        \label{fig:missdata12}
         \vspace{2em}
    \end{minipage}\hspace{0.5em}
    \begin{minipage}{0.45\textwidth}
        \centering
        \includegraphics[width=\linewidth]{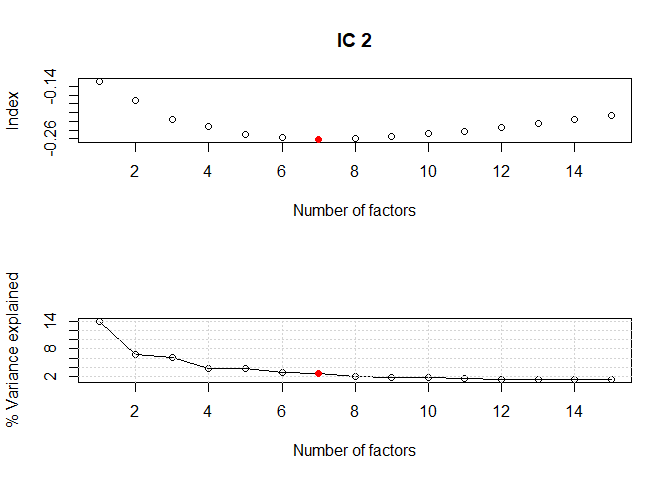}
        \caption{Output of \emph{tuneFactors()} showing the best number of factors to use for the exports data set. }
        \label{fig:tunefactorsexports}
        \vspace{2em}
    \end{minipage}
\end{figure}

\section{Code for the Pseudo Real-Time Nowcasting Exercise of Section 4.2}\label{appendixG}
The expanding window nowcasting loop from August 2018 to July 2022 can be implemented in R using:
\begin{verbatim}
## Start and end dates 
start_date <- as.Date("2018/08/01")
end_date = as.Date("2022/07/01")

## Generating range of dates
range <- seq(start_date, end_date,"months")

## Ragged edge variable information (match to column variables)
# 98 series with 2 month lag (targets,IoP)
# 333 series with 1 month lag (CPI,PPI,exchange,conf)
# 14 series with 0 month lag (google trends)
ragged_edge_values = c(rep(2,98), rep(1,333), rep(0,14))

## Remove missing jan 2004 and make time series from Feb 2004
new_data = new_data[-1,]
new_data = ts(new_data, start = c(2004,2), frequency = 12)

## Expanding window nowcasting loop and store the DFM and SDFM fits
fit.dfm <- fit.sdfm <- list()

for(t in 1:length(range)){
  
  date = range[t]
  date.year = as.numeric(format(date, "%Y"))
  date.month = as.numeric(format(date, "%m"))
  
  Y = window(new_data, end = c(date.year, date.month))
  Y = raggedEdge(Y, ragged_edge_values)
  
  fit.dfm[[t]] <- sparseDFM(Y, r = 4, alg = 'EM')
  fit.sdfm[[t]] <- sparseDFM(Y, r = 4, q = 9, alg = 'EM-sparse')
  
}
\end{verbatim}
Using the DFM fits, we can extract nowcasts
and then obtain the mean absolute error values in Table \ref{tab:pseudo} using:
\begin{verbatim}
## Get the original data in levels for the targets (with Jan 2004 removed)

data_exports = data[-1,]
data_exports = data_exports[,1:9]

## Compute errors for the fits at horizon 1 and 2 for each nowcasting window

# n = 175 represents August 2018 (for the 1st nowcasting window)
n = 175

err1.dfm = err2.dfm = err1.sdfm = err2.sdfm = c()

for(t in 1:48){
  
  # extract nowcasts for horizon 1 and 2 of the targets 
  hor1.dfm = fit.dfm[[t]]$data$fitted.unscaled[n-1,1:9]
  hor2.dfm = fit.dfm[[t]]$data$fitted.unscaled[n,1:9]
  hor1.sdfm = fit.sdfm[[t]]$data$fitted.unscaled[n-1,1:9]
  hor2.sdfm = fit.sdfm[[t]]$data$fitted.unscaled[n,1:9]
  
  # true data 
  data.obs = data_exports[n-2,]
  
  # nowcasts 
  for1.dfm = data.obs + hor1.dfm
  for2.dfm = for1.dfm + hor2.dfm
  for1.sdfm = data.obs + hor1.sdfm 
  for2.sdfm = for1.sdfm + hor2.sdfm
  
  # mean absolute error
  err1.dfm[t] = mean(abs(for1.dfm - data_exports[n-1,]))
  err2.dfm[t] = mean(abs(for2.dfm - data_exports[n,]))
  err1.sdfm[t] = mean(abs(for1.sdfm - data_exports[n-1,]))
  err2.sdfm[t] = mean(abs(for2.sdfm - data_exports[n,]))
  
  # move window along a month 
  n = n + 1 
}

## Print out the results 
# DFM horizon 1 
quantile(err1.dfm)
#>        0%       25%       50%       75%      100% 
#>   80.7734  152.4034  232.8558  455.0769 1990.3548

# Sparse DFM horizon 1
quantile(err1.sdfm)
#>         0%        25%        50%        75%       100% 
#>   82.47065  141.11771  212.39231  306.92548 2030.41868

# DFM horizon 2
quantile(err2.dfm)
#>         0%        25%        50%        75%       100% 
#>   82.87274  201.82086  302.67172  515.08069 2162.27714

# Sparse DFM horizon 2
quantile(err2.sdfm)
#>        0%       25%       50%       75%      100% 
#>   84.0062  161.9366  233.7898  366.3923 2129.1993

## Average 

# DFM horizon 1 
mean(err1.dfm)
#> [1] 373.72

# Sparse DFM horizon 1
mean(err1.sdfm)
#> [1] 297.2334

# DFM horizon 2
mean(err2.dfm)
#> [1] 437.1499

# Sparse DFM horizon 2
mean(err2.sdfm)
#> [1] 357.6024
\end{verbatim}

\section{Timing Comparison Between Univariate and Multivariate KFS Equations}\label{appendixH}
It is recommended to use \verb+kalman = "univariate"+ when setting \verb+err = "IID"+ and \verb+kalman = "multivariate"+ when \verb+err = "AR1"+ for the fastest results. In Section 2.2, it was shown how the univariate treatment of the classic multivariate KFS equations \citep{koopman2000fast} can lead to substantial computational gains over the classic multivariate approach \citep{shumway1982approach} since matrix inversions become scalar divisions. The only matrix inversion that is required is in the update of the lagged cross-covariance matrix $\bP_{t-1,t|n}$ that uses \cite{jong1988covariances} theorem:
\begin{equation*}
\bP_{t,t-1|n} = \bP_{t|n}(\bP_{t|t-1})^{-1}\bA \bP_{t-1|t-1}.
\end{equation*}
The matrix needing to be inverted, $\bP_{t|t-1}$, is of dimension equal to the number of states. When idiosyncratic errors are assumed to be IID, the number of states is just equal to the number of factors $r$, a small number, and hence matrix inversion is rapid. However, in the AR(1) case, the errors augment the state variable and the number of states becomes $p + r$; leading to a matrix that is much slower to invert. 

\begin{figure}[h!]
    \centering
    \includegraphics[width=0.49\linewidth]{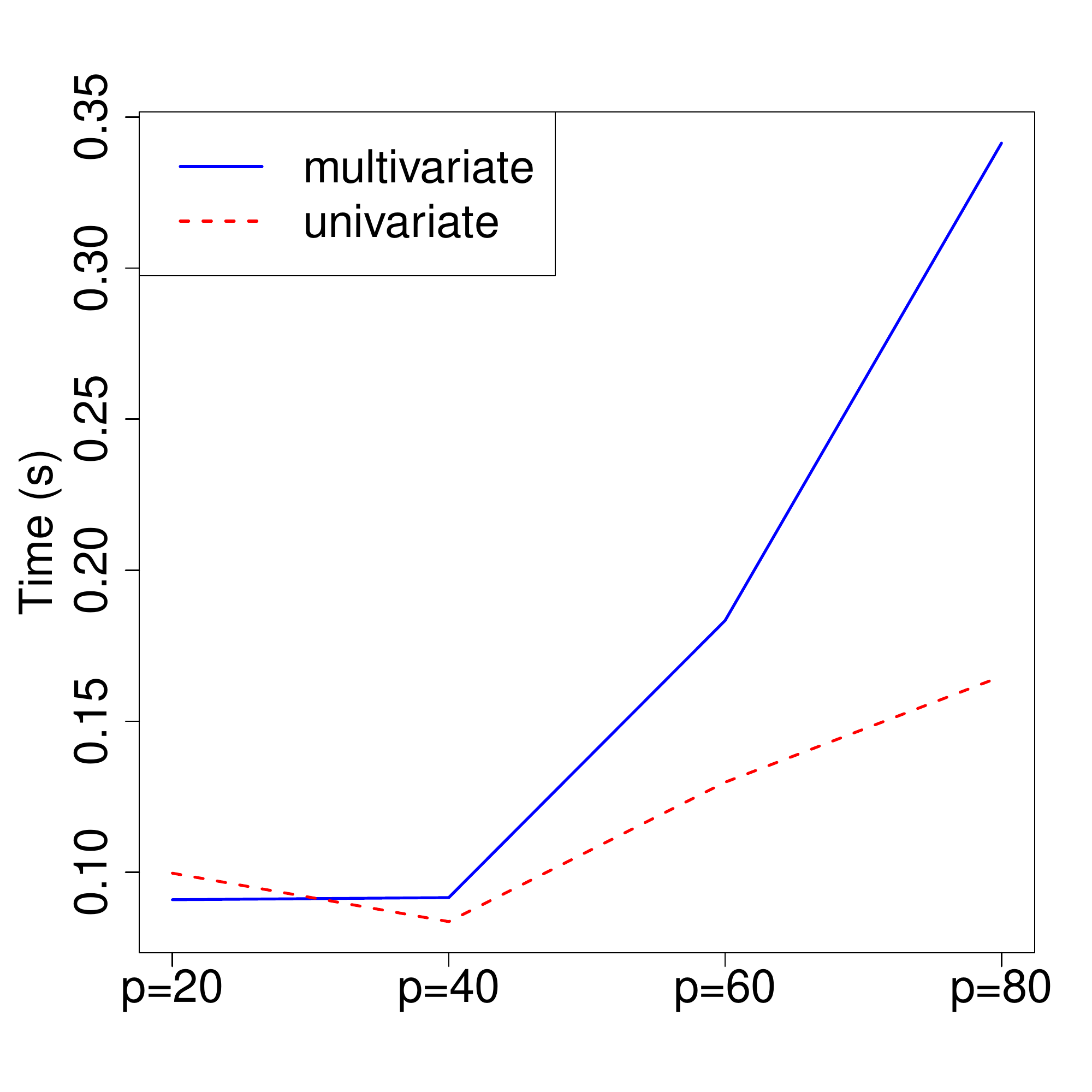}
    \includegraphics[width=0.49\linewidth]{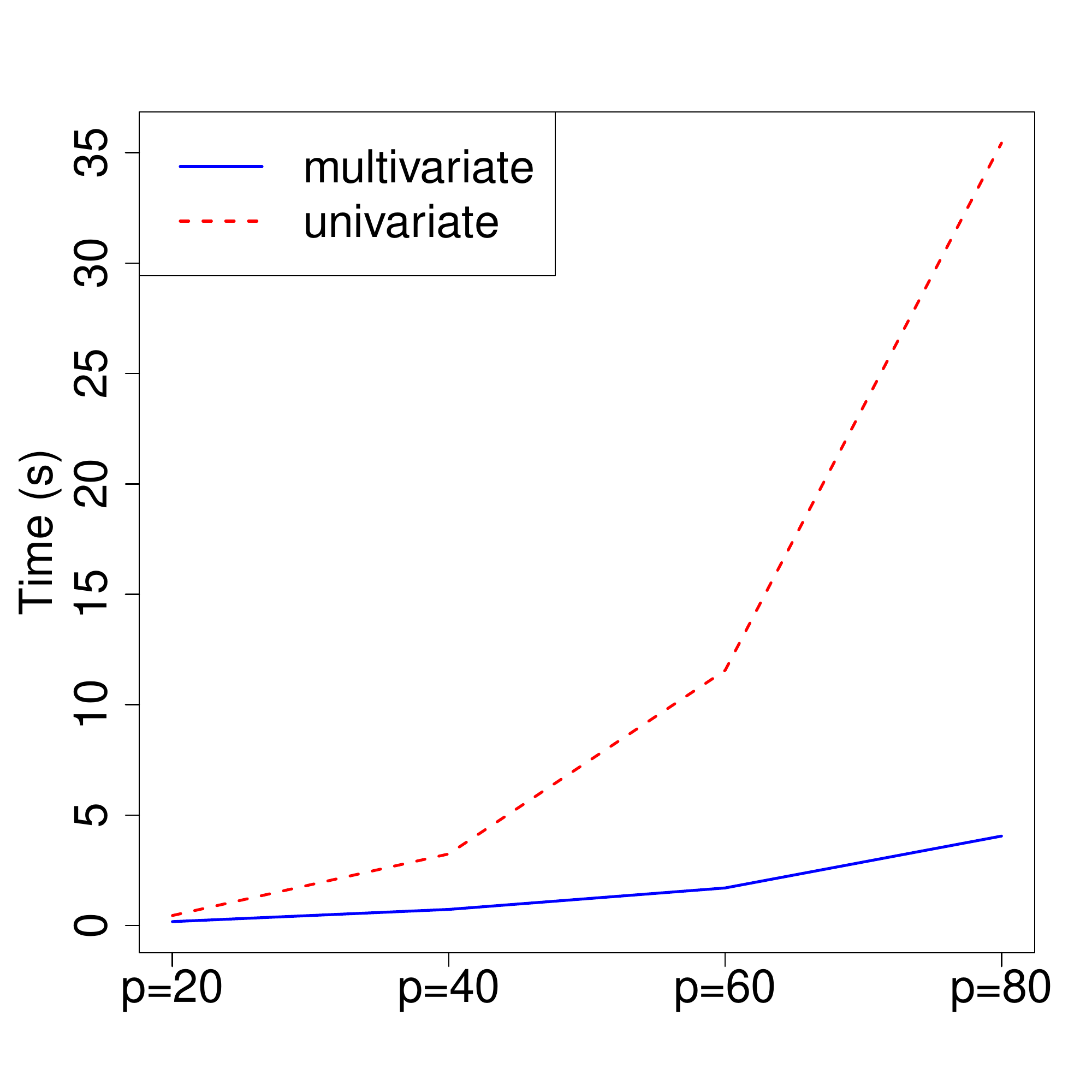}
    \caption{Average time from 10 experiments the EM algorithm (setting \emph{alg = "EM"}) takes with the multivariate and univariate KFS methods with \emph{err = "IID"} on the left and \emph{err = "AR1"} on the right.}
    \label{fig:kfstimes}
\end{figure}

To test the timing comparisons we simulate data from a DFM with $\blam = N(\bm{0}_p, \bI_p)$, $\bsig_\epsilon = \bI_p$, $\bA=0.8 \times \bI_r$ and $\bsig_u = (1-0.8^2) \times \bI_r$ using 2 factors and fit a DFM using \verb+alg = "EM"+ \citep{banbura2014maximum} comparing the multivariate and univariate KFS methods when errors are assumed to be IID and AR(1). Figure \ref{fig:kfstimes} shows the average time from 10 experiments with fixed number of observations, $n=100$, and varying dimension $p$, with the \verb+err = "IID"+ case on the left and the \verb+err = "AR1"+ case on the right. The IID case shows the clear computational gain of the univariate filtering KFS approach, and this becomes even more prominent for larger $p$. The AR(1) setting clearly takes a longer amount of time to run and the univariate filtering is slower than the classic multivariate approach. 

Due to the significant time difference between assuming IID errors compared with AR(1) errors, it is recommended to assume IID errors and then perform residual analysis afterwards for forecasting purposes. 

\end{document}